\documentclass[11pt,a4paper]{article}
\pdfoutput=1
\usepackage{jheppub}
\usepackage{amsfonts}
\usepackage{bbm}
\usepackage{graphicx}
\usepackage{caption}
\usepackage{subcaption}
\usepackage{bigints}
\usepackage[normalem]{ulem}
\usepackage{slashed}

\DeclareMathOperator{\sech}{sech}

\def\a{\alpha}

\def\c{\gamma} \def\g{\gamma}

\def\h{\eta}

\def\l{\lambda}    
\def\m{\mu}
\def\n{\nu}

\def\th{\theta}

\def\x{\xi}

\def\D{\Delta}

\def\h{\eta}

\def\L{\Lambda}

\def\X{\Xi}

\newcommand{\ndt}{\noindent}

\newcommand{\nn}{\nonumber}

\def\p{\partial}

\def\bea{\begin{eqnarray}}
\def\eea{\end{eqnarray}}
\def\be{\begin{equation}}
\def\ee{\end{equation}}
\def\ba{\begin{align}}
\def\ea{\end{align}}

\newcommand{\bem}{\begin{pmatrix}}
\newcommand{\eem}{\end{pmatrix}}

\def\={\;  = \;}
\def\+{\, + \,}

\def\wt{\widetilde}

\def\bar{\overline}

\def\rt2{\sqrt{2}}

\title{Boundary conditions and localization on $AdS$: Part 2  General analysis}

\author[a]{\small{Justin R.  David},}
\author[b,d]{\small{Edi Gava},}
\author[c]{\small{Rajesh Kumar Gupta},}
\author[d]{\small{Kumar Narain}}
\affiliation[a]{\small{Centre for High Energy Physics, Indian Institute of Science,\\
C. V. Raman Avenue, Bangalore 560012, India.}}
\affiliation[b]{\small{INFN, sezione di Trieste, Italy}}
\affiliation[c]{\small{Department of Mathematics, King's College London, The Strand, London WC2R 2LS, UK}}
\affiliation[d]{\small{ICTP, Strada Costiera 11, 34151 Trieste, Italy}}

\emailAdd{ justin@cts.iisc.ernet.in}
\emailAdd{gava@ictp.it}
\emailAdd{rajesh.gupta@kcl.ac.uk}
\emailAdd{narain@ictp.it}
\abstract{
We develop the method of Green's function to evaluate the one loop 
determinants that arise in localization of supersymmetric field theories on 
$AdS$ spaces.  The theories we study have at least ${\cal N}=2$ supersymmetry and 
normalisable boundary conditions are consistent with supersymmetry. 
We then show 
 that under general assumptions 
 the  variation of the  one loop determinant  with respect to the 
 localizing background   reduces to a total derivative.  Therefore it 
 receives 
contributions only from the origin of $AdS$ and from asymptotic infinity. 
From expanding both the Greens function and the quadratic operators 
at the origin of $AdS$ and asymptotic infinity, we show that the variation of the 
one loop determinant is proportional to an integer. 
Furthermore, we show that this integer is an index of a first order differential operator.
We demonstrate  that these assumptions are valid for Chern-Simons theories 
coupled to chiral multiplets on $AdS_2\times S^1$. 
Finally we use our results to show  that $U(N_c)$ Chern-Simons theory  at level $k$ coupled
to $N_f$ chiral multiplets and $N_f$ anti-chiral multiplets in the fundamental 
obeys level-rank duality on $AdS_2\times S^1$. 
}

\begin{document}

\maketitle
\section{Introduction}

Supersymmetric localization on compact spaces and its applications  has been studied extensively 
recently, see \cite{Pestun:2016zxk} for a  recent review. 
This area  began with the work of Witten \cite{Witten:1992xu}  and was 
developed in the works of 
\cite{Nekrasov:2003af,Nekrasov:2003rj,Pestun:2007rz} 
to enable the evaluation of observables in supersymmetric quantum field theories. 
The exact computation of supersymmetric  partition functions and Wilson lines 
served as highly non-trivial checks of AdS/CFT \cite{Marino:2009jd,Benini:2016rke,Benini:2015eyy}. Field theories defined on a compact 
space serve as standard examples  for applying the method of localization. 
This is because the method relies on identifying a  fermionic symmetry $Q$. 
The Lagrangian including the localizing term is symmetric under $Q$ only upto boundary 
terms and restricting the space to be compact ensures that these boundary terms do not arise. 

The systematic extension of the method of supersymmetric localization is an important
 problem.  Non-compact spaces which  form the canonical examples to study localization  are of the 
 form $AdS_n\times S^m$.  This is mainly due to the variety of applications of supersymmetric 
 theories  on such spaces. For example, 
  localization of ${\cal N}=2$ gravity on $AdS_2\times S^2$ is 
 important for obtaining the exact entropy of BPS black holes in these theories
  \cite{Dabholkar:2010uh,Dabholkar:2011ec,Gupta:2012cy,Dabholkar:2014ema,Murthy:2015yfa,Gupta:2015gga}. 
 Similarly the exact evaluation of the   supersymmetric partition function of ${\cal N}=8$ supergravity 
 on $AdS_4$ 
 serves as an important check of the holographic duality with ABJM theory 
 \cite{Dabholkar:2014wpa,Cabo-Bizet:2017jsl}. 
 
As demonstrated in~\cite{David:2016onq},  when the method of supersymmetric localization is applied to 
non compact spaces  one needs
to carefully examine the  boundary conditions implemented on the fields. 
The boundary conditions of both the bosonic and fermionic fields must be chosen so that 
they are consistent with the superysmmetric transformations.  They also must be chosen 
so that boundary terms that arise under the action of $Q$ on both the original action as 
well as the localizing term vanish. Furthermore, the path integral must be well defined 
under these boundary conditions. Normalizable boundary conditions on all fields 
ensure that the boundary terms at asymptotic infinity vanish as well as the path integral is 
well defined. However normalizable boundary conditions may not always be compatible 
with supersymmetry. 

In \cite{David:2018pex}, the method of Greens function was introduced to evaluate one loop determinants 
that arise in localization.  This was done for the ${\cal N}=2$  chiral multiplet on $AdS_2\times S^1$. 
The method involved studying the variation of the one loop determinant under 
a parameter $\alpha$ \footnote{$\alpha$ parametrises the vector multiplet background.}
 that parametrises the localizing background 
  and then integrating 
with respect to $\alpha$.  
It  was shown that whenever normalizable boundary conditions also are consistent 
with supersymmetric transformations, the  variation of the one loop determinant 
reduces to a total derivative and one only needs to evaluate the boundary contributions 
from the origin of $AdS_2$ and the asymptotic infinity.  
Furthermore, it was demonstrated that the final result for the one loop determinant agrees with the
index  whenever the boundary conditions are normalizable and supersymmetric.  

In this paper we develop the Green's function method further. 
A brief outline of the Greens function method is the following. 
Let ${\cal D}_B(\alpha) $ be the bosonic operator and ${\cal D}_F(\alpha) $ be the fermionic 
operator that occurs in the evaluation of the one loop determinants. 
They depend on the classical localising background through the parameter $\alpha$. 
Then the variation of the one loop determinants with respect to $\alpha$ is given by 
\begin{equation}\label{var}
\frac{\delta}{\delta \alpha} \ln Z_{\rm 1-loop} (\a)
= {\rm Tr} [ G_F \frac{\delta}{\delta \alpha} {\cal D}_F(\alpha) ]
- \frac{1}{2} {\rm Tr} [ G_B \frac{\delta}{\delta \alpha} {\cal D}_B(\alpha) ]\,,
\end{equation}
Here $G_B, G_F$ are the bosonic and fermionic Greens function corresponding to the 
operator ${\cal D}_B$ and ${\cal D}_F$ respectively. 

We show that under some general assumptions which hold for theories  with at least 
${\cal N}=2$ supersymmetry on 
spaces of the form $AdS_n\times S^m$ the variation of the one loop determinant  with respect 
to  $\alpha$ which parametrises the  localising backround
always reduces to a total derivative.  
This reduction to a total derivative holds whenever supersymmetric boundary conditions 
are compatible with normalisable boundary conditions. 
The general assumptions that we make relate to the properties  of the second order operators, 
${\cal D}_B$ and ${\cal D}_F$,
that arise in these theories in the evaluation of the one loop determinants. 
These assumptions enable the evaluation of the variation given in (\ref{var}). 
Then integrating with respect to $\alpha$ we can obtain the one loop determinant. 
In this paper we demonstrate that these properties hold 
for both the vector multiplet as well as the chiral multiplet on $AdS_2\times S^1$. 
We have also verified that it continues to hold for the vector as well as the hypermultiplet on 
$AdS_2\times S^2$ \cite{dggn}.
We suspect that the general assumptions are properties that hold whenever the actions 
have at least ${\cal N}=2$ supersymmetry but at present we do not have a proof.

Here we state these assumptions in a qualitative form. In  the next section 
we make these quantitative.  These assumptions are made on the second order 
differential operators that appear after one reduces the operators ${\cal D}_B, {\cal D}_F$  to 
only the radial equation  equation parametrising the $AdS$ direction. 
This reduction is made  by expanding all the fields in an appropriate basis. 
For example, it is 
 the Fourier basis corresponding to the two  $S^1$'s for the case of  $AdS_2\times S^1$ 
 \begin{enumerate}
\item The  matrix second order  operator corresponding to the 
one loop bosonic determinant reduces to a certain block diagonal form
in a special gauge.  The operators are Hermitian and non-degenerate and have 
regular singularities at the origin of $AdS$ and the boundary. 
This last assumption enables a Forbenius series expansion of the solutions 
at these points. 
\item The matrix second order operator corresponding to the one loop 
fermionic determinants also reduce to a certain block form. 
All the  second derivatives  occur  only with  terms involving the ghosts.
The operator is Hermitian.  It is only certain components of the 
block form that contain the dependence on $\alpha$ which parametrises the 
background. 
 \item The bosonic operator and the fermionic operators are 
 related to each other by factors of $Q^2$.  This follows from supersymmetry. 
 Therefore the fermionic 
 solutions can be found in terms of the bosonic ones.  
 \item  The Greens function for the bosonic operator exists and this 
 in turn implies the Greens function for the fermionic operator 
 can be constructed from that of the bosonic Greens function.  
\end{enumerate}
Using these assumptions it can be shown that the variation in (\ref{var}) reduces to 
a total derivative. Therefore, the behaviour of the Greens functions as well as the second order operators at the origin of $AdS$ and at infinity 
determine the variation (\ref{var}).  The result for the variation  is given in equation \eqref{boundaryterm.1}. 
Then finally using  assumptions of the behaviour  of certain components of the 
fermionic matrix operator at these points, the variation can be evaluated. 

Our main result is that we show that the variation of the one loop determinant given in 
 (\ref{var})   is an  integer times the variation of $\frac{1}{2} \ln( Q^2) $. 
The integer is determined by  the difference  between the  
number of  allowed solutions to a first order 
differential equation that occurs from 
the fermionic operator  at the origin 
and at asymptotic infinity of the $AdS$. 
 The result is given in equation (\ref{main_result}). 
Thus, the final result for the one loop determinant resembles an index of an operator. We then identify this operator and show that the one loop determinant is expressed in terms of index of this operator.

As we mentioned earlier we verify that these assumptions hold for the case of the ${\cal N}=2$ 
vector as well as the chiral multiplet on $AdS_2 \times S^1$. 
We also show that normalisable boundary conditions imply supersymmetric boundary conditions
for the vector multiplet  provided $L^2 > \frac{3}{4}$. 
 Here $L$ is the  ratio of the radius of $AdS_2$ to 
$S^1$. 
For the chiral multiplet of R-charge $\Delta$  the conditions that ensure normalisable boundary conditions 
are also supersymmetric is that there should be no integer $n$ in the interval~$( \frac{\Delta- 1}{2L} , \frac{\Delta}{2L})$ was obtained in~\cite{David:2018pex}. 

We apply our results  to ${\cal N}=2$ Chern Simons with  $N_f$ chiral multiplets in the fundamental 
and $N_f$ anti-chiral multiplets in the fundamental and show that the partition function of the  theory with 
gauge group $U(N_c)$ at level $k$  is identical  to the theory with the gauge group $U(|k| + N_f - N_c)$ at 
level $-k$ and with the same matter content. 
That is level-rank duality continues to hold 
when the theory is placed on $AdS_2\times S^1$.

It is important to mention that our gauge fixing condition is a generalisation of the covariant 
gauge condition which is given by 
\begin{equation} \label{oldg}
\cosh^2 r \nabla^{\hat\mu} (  \frac{ 1}{\cosh^2r } a_{\hat\mu}  )  + \partial_t a_t  = 0\,. 
\end{equation}
This gauge condition was first used in \cite{David:2016onq}.  Here $r$ is the radial coordinate in $AdS_2$,  $\hat\mu$  refers to the  two coordinates on $AdS_2$ and $t$ refers to the coordinate on $S^1$. 
This gauge choice ensures that the operators that occur the operators that occur in the analysis of the Greens function of the bosons is block diagonal.  We have seen that the results are independent of 
gauge choice. 
We show in  appendix  \ref{appenC} that for the bosonic  $U(1)$  Chern-Simons theory,
the partition function evaluated in 
a one parameter set of gauge conditions that interpolate between the covariant gauge 
and the condition  in (\ref{oldg}) remains the same.

This paper is organised as follows. 
In  section  \ref{genproof} we present the details of the assumptions made on the properties 
of  the quadratic operators that appear in localization of  at least ${\cal N} = 2$ theories  on 
$AdS$ spaces.  In section \ref{calcbterm} we make further assumptions on 
the behaviour of the terms in the fermionic kinetic term at the boundary of $AdS$ and at the 
origin.   
We then present our proof that under these assumptions 
the variation of the one loop determinant is an integer times the variation of 
$\frac{1}{2} \ln (Q^2 )$ is given in section  \ref{calcbterm}. In section \ref{IndexofC}, we show that this integer is the index of a first order matrix differential operator appearing in the fermionic kinetic term.
In section  \ref{chern-section}  we introduce ${\cal N}=2$ Chern Simons  theory on $AdS_2\times S^1$,  the 
localizing term as well as the gauge fixing condition. 
We also determine the behaviour of all fields at asymptotic infinity of $AdS_2$ so that 
they are all normalisable.
In section \ref{eom-section}, we demonstrate that  the  general assumptions made on the properties of the 
second operators that occur in evaluating one loop determinants 
 in section  hold for the case of ${\cal N}=2$ Chern-Simons theory on $AdS_2\times S^1$. 
 We also derive the conditions under which  normalizable boundary conditions are consistent with supersymmetry.  Finally we obtain the variation of the one loop determinant and 
 demonstrate that it is an integer times the variation of $\frac{1}{2} \ln (Q^2)$. 
 We show that the result agrees with that obtained in \cite{David:2016onq}. 
 In section \ref{s-duality} we apply our analysis to evaluate the supersymmetric partition function 
 of $U(N_c)$ Chern-Simons theory  on $AdS_2\times S^1$ 
 coupled with $N_f$ chiral multiplets  in the fundamental 
 and an equal number of chiral multiplets in the anti-fundamental. 
 From the expression of the  partition function we  demonstrate  this theory obeys  level-rank duality. 
 Section \ref{discuss} contains our conclusions. 
 Appendix \ref{appenA} and \ref{appenB} provide the details of the supersymmetic variations as well as 
 the equations of motion of all the fields about the localization  background. 
 Appendix \ref{appenC} contains the evaluation of the partition function of $U(1)$ Chern-Simons theory 
 in a one parameter set of gauge conditions which interpolate between 
 the covariant gauge and  the gauge in (\ref{oldg}).

\section{A general proof} \label{genproof}
In this section, we will present a general discussion about the one loop computations in supersymmetric localization on a general manifold for vector and matter multiplets. Our discussion will be based on the Green's function method which was used in~\cite{David:2018pex} to compute the path integral of chiral multiplet on AdS$_{2}\times$S$^{1}$. In the computation of path integral using the supersymmetric localization technique, we need to compute the one loop determinant of the operators about the localization background. In the Green's function approach, developed in~\cite{David:2018pex}, we computed the variation of the one loop determinant instead i.e.
\begin{equation}\label{greendet.1}
\frac{\delta}{\delta \alpha} \ln Z_{\rm 1-loop} (\a)
\= {\rm Tr'} [ G_F \frac{\delta}{\delta \alpha} {\cal D}_F(\alpha) ]
- \frac{1}{2} {\rm Tr'} [ G_B \frac{\delta}{\delta \alpha} {\cal D}_B(\alpha) ]\,,
\end{equation}
where~${\cal D}_F(\alpha)$ and~${\cal D}_B(\alpha)$ are fermionic and bosonic kinetic operator, respectively and $G_F$ and $G_B$ are their Green's functions. Also, $\a$ is some parameter which enters in both bosonic and fermionic differential operator and the ``$\text{Tr}$'' in \eqref{greendet.1} is the space-time as well as matrix trace over non zero modes. Typically, we choose this parameter to the one which parametrises the localization background. The one loop determinant, up to a constant in $\a$, is then obtained by integrating the right hand side of \eqref{greendet.1} with respect to $\a$. 

The choice of the parameter $\a$ is arbitrary as it was shown in~\cite{David:2018pex}, the final result of the one loop determinant is independent of the choice of the parameter with respect to which we decide to vary the one loop determinant. Thus in this method, we need compute the Green's function of the differential operator which appears in the one loop computations. One of the remarkable simplifications occur in this approach is that when the boundary conditions of the fields are consistent with supersymmetry, the variation~\eqref{greendet.1} is a total derivative and contributions to one loop determinant comes from the boundary behaviour of the solutions of Equations of motion of all the fields in the chiral multiplet. We find that this is quite generic feature of the supersymmetric localization and independent of the multiplet and spaces i.e. if the boundary conditions of the fields are consistent with supersymmetry, the variation is always a total derivative,

Our method presented below is quite generic and, in particular, very useful for the localization computation in non compact spaces such as AdS space which also involve imposing a boundary conditions. We start with stating the notation and the set up.

\ndt{\bf Set up:}
\begin{enumerate}
\item In the vector multiplet fields, after integrating out the auxiliary fields as well as $b$ (BRST partner of the ghost $\wt c$) ghost we are left with the vector field, scalar fields, ghost field $c$ and fermions. We denote the bosonic fields by $X_{0}$ and $\sigma$, where $\sigma$ is the scalar field which parametrises the localization manifold. The bosonic field $X_{0}$ is a $(k+1)$ component column vector.
 In the case of $AdS_2\times S^1$, we have
$k +1=3$, $X_0$ consists of the gauge field $a_{\hat\mu}, a_t$. 
Since the method always requires a scalar which takes a non zero value on the localization manifold, 
the method is suitable for theories with at least ${\cal N}=2$ supersymmetry. 
\item The fermionic fields are grouped as $QX_{0}$ and $(c,X'_{1})$. The fermionic field $QX_{0}$ and $X'_{1}$ are $(k+1)$ and $k$ component column vector, respectively.
\item In the matter multiplet fields, after integrating out the auxiliary fields we are left with scalar fields which we denote by $X_{0}$ and the fermionic fields are decomposed as $QX_{0}$ and $X_{1}$. We assume that the scalar fields in the matter multiplet do not acquire non zero value on the localization manifold.
\end{enumerate}
With this set up, our method of localization computations will be based on the following assumptions:

\ndt{\bf Assumptions:}
\begin{enumerate}
\item Fields are functions of a non periodic coordinate $r$. In particular, it is assumed that we have done Kaluza Klein reduction in the rest of the coordinate and the Lagrangian for each KK mode is a function in one variable $r$. We will take the range of $r$ to be from $0$ to $\infty$ for convenience (precise interval is not important for most of the presentation).
\item For the vector multiplet calculations, we need to add gauge fixing functional $\mathcal G(A)$ in the path integral. We assume that  the gauge fixing condition ${\cal{G}}(A)$ is such that after eliminating auxiliaries and $b$,  the bosonic equation for $\sigma$ decouples from the rest of the bosonic fields $X_{0}$ \footnote{For the vector multiplet case we add the following gauge fixing term in the $QV$ action:
 $Q\text {tr}(\tilde{c} ( {\cal{G}} + \xi b ))$
where $\xi$ is a parameter. It turns out that in order to decouple the Equations of motion of $\sigma$ field from rest of the bosonic fields, one needs to add a $Q$-exact term to the localizing action of the form $Q(\text{ tr}( \alpha [c, {\cal{G}}]))$ , where $\a$ is some constant (in general it is related to localization background), which can also be thought of as redefining $\tilde{c}\rightarrow \tilde{c}+ [\alpha ,c]$.}. This choice is not necessary but it will simplify some of the calculations. The bosonic equations can, therefore, be written as the matrix operator
\begin{equation}
\begin{pmatrix}
 A_1^b(r)& 0\\
  0& A_2^b(r)
\end{pmatrix}
\begin{pmatrix} X_0\\
  \sigma
 \end{pmatrix}\equiv M_b \begin{pmatrix}
 X_0\\
  \sigma
 \end{pmatrix}
 \end{equation}
For the matter multiplet case there is no second block corresponding to $\sigma$. 
\item We assume that $M_b$ is hermitian second order matrix differential operator:
  \begin{equation}
   M_b(r)\= M_b^{(2)}(r) \frac{d^2}{dr^2}+  M_b^{(1)}(r) \frac{d}{dr}+  M_b^{(0)}(r)
  \end{equation}
  where $M_b^{(2)}(r)$, $M_b^{(0)}(r)$ and $M_b^{(0)}(r)$ are $(k+2) \times (k+2)$ matrices and $M_b^{(2)}(r)$ is non-degenerate. It implies that $A_1^b(r)$ and $A_2^b(r)$ are hermitian and second order matrix differential operators
and the coefficient of $\frac{d^2}{dr^2}$ is non-degenerate for all $r \in (0, \infty)$. At $r=0$ and $u=e^{-r}=0$ (i.e. the two boundaries of the one-dimension problem) and the operators  $A_1^b$ and $A_2^b$ have regular singularities.\item For the fermionic fields the equations are:
 \begin{equation}
\begin{pmatrix}
 A_{11}(r)& A_{12}(r)&B(r)\\
  A_{21}(r)&A_{22}(r) &0\\
  C(r)&0&D(r)
 \end{pmatrix}\begin{pmatrix}
 QX_0\\
  c\\X_1'
 \end{pmatrix} \equiv M_f(r) \begin{pmatrix}
 QX_0\\
  c\\X_1'
  \end{pmatrix} 
 \end{equation}
 Here, generically, $A_{11}(r), A_{12}(r), A_{21}(r)$ and $A_{22}(r)$ are $(k+1)\times(k+1), (k+1)\times 1, 1\times (k+1)$ and $1\times 1$ matrix differential operators, respectively. Similarly, $B(r),C(r)$ and $D(r)$ are $(k+1)\times k,k\times (k+1)$ and $k\times k$ matrix differential operators, respectively.
 \item $M_f(r)$ is assumed to be Hermitian.
In particular, this means that $A_{11}(r), A_{22}(r)$ and $D(r)$ are Hermitian while $A(r)^{\dagger}_{21}\=A_{12}(r)$ and 
$ B(r)\=C(r)^{\dagger}$.
\item $D(r)$ is purely algebraic and is invertible $k\times k$ matrix.
 $A_{11}(r)$, $B(r)$ and $C(r)$ involve only first order differential operators. The only two derivative term in the localizing action are the ones that involve ghost $c$. What this
means is that $A_{21}(r)$, $A_{12}(r)$ and $A_{22}(r)$ involve second order differential 
operators. 
\item Requiring that the action is supersymmetric implies that one can obtain the Equations of motions for fermionic fields from those of the bosonic fields upto a factor of $Q^{2}$. This implies that there exist a matrix first order differential operator $E$ and its adjoint $E^{\dagger}$ such that
\begin{equation}\label{SimilarityTransf.1}
 \hat M(r)\equiv E(r)^{\dagger}M_f(r)E(r) \= \begin{pmatrix} 
 \gamma_1 A_1^b(r)& 0&0\\
0&\gamma_2 A_2^b(r)\\
  0&0&D(r)
 \end{pmatrix} 
\end{equation}
 It is not very hard to find $E(r)$ which does the above and is given by
\be
  E(r)\=  \begin{pmatrix} 
 1& 0&0\\
  K& f(r)&0\\
  -D(r)^{-1}C(r)&0&1
 \end{pmatrix} 
\ee
and for this choice of $E(r)$, the constants are $\g_{1}\=\frac{1}{Q^{2}}$ and $\g_{2}\=Q^{2}$. Here $K$ is a~$(k+1)$-component row vector and $f(r)$ is a scalar function which is independent of the parameter $\a$. More explicitly, the relations are\footnote{The similarity transformations~\eqref{SimilarityTransf.1} are obtained by following supersymmetry which implies that the Equations of motion for $X_{0}$ and $QX_{0}$ are identical upto a factor of $Q^{2}$. Similarly, the Equation of motion for the ghost field $c$ is also related to $\sigma$. This relations follows because $Qc=f(r)\sigma+k\cdot X_{0}$ where $k$ is some vector which is usually related to the killing vector. If we define the field $c'$ as $c'=f(r)^{-1}\Big(c-\frac{1}{Q^{2}}k\cdot QX_{0} \Big)$, then we see that supersymmetry implies the Equation of motion for $c'$ is identical to that of $\sigma$ upto a factor of $Q^{2}$.}
\bea\label{RelationAij}
&&A_{12}(r)\=-K^{\dagger}A_{22}(r),\quad A_{21}(r)\=-A_{22}(r)K\,\nn\\
&&A_{11}(r)-K^{\dagger}A_{22}(r)K-B(r)D(r)^{-1}C(r)\=\g_{1}A^{b}_{1}(r),\quad f(r)^{\dagger}A_{22}(r)f(r)\=\g_{2}A^{b}_{2}(r)\nn\\
\eea

\item The Greens fn for $A_1^b$ exists. This means that $A_1^b$ has  no zero modes. The differential operator $A_2^b$ can have zero modes. Typically, these correspond to the variation of the 
saddle point, which happens only for modes that are constant along space orthogonal to $AdS_2$ and  for which we 
already have collective coordinate integration. This case will be discussed separately. 
\end{enumerate}
\subsection{Green's function}\label{Sec.greenfn.}
In this section, we will construct the Green's function for both the fermionic and bosonic kinetic operators and discuss the relation between the two. We will find that the fermionic Green's function can always be constructed from the bosonic Green's function provided their boundary conditions agree with supersymmetry. 

We start with the bosonic Green's function. The bosonic Green's function satisfies the equation
\be
M_{b}(r)G_{b}(r,r')\=\delta(r,r')\,I_{k+2}\,.
\ee
Here $I_{k+2}$ is $(k+2)$-dimensional identity matrix. In general, the differential operator $M_{b}(r)$ could have zero modes. Since, in the path integral we integrate over only non zero modes, therefore, we are interested in computing only the Green's function for the non zero modes.

Let the solution for the Green's function equation for $r < r'$ be 
\begin{equation}\label{bosonicGreenfn.1}
G^{<}_b(r,r')\=\begin{pmatrix} 
 G_1(r,r')& 0\\
  0& G_2(r,r')
\end{pmatrix} \,,
 \end{equation}
 and for $r > r'$ be
 \begin{equation}\label{bosonicGreenfn.2}
G^{>}_b(r,r')\=\begin{pmatrix} 
 G'_1(r,r')& 0\\
  0& G'_2(r,r')
\end{pmatrix} \,.
 \end{equation}
 Furthermore, $G^{<}_b(r,r')$ is smooth at $r\=0$ and satisfy the allowed boundary conditions 
 at $r' \= \infty$
 while $G^{>}_b(r,r')$ is smooth at $r'\=0$ and satisfy the allowed boundary conditions at $r \= \infty$. It is important to note that these boundary conditions on the Green's function are exactly the same boundary condition which impose on the bosonic fields.
 
Since, $M_{b}(r)$ is a 2nd order differential operator, these Green's function also satisfy the continuity/discontinuity relations:
 \begin{eqnarray}
 \lim_{r'\rightarrow r}(G^{>}_b(r,r')-G^{<}_b(r,r'))&=&0\,,\\
 \lim_{r'\rightarrow r}\partial_r (G^{>}_b(r,r')-G^{<}_b(r,r'))&=&( M^{(2)}_b(r))^{-1} \,.
 \label{disc}
\end{eqnarray}
 Next, we will determine the fermionic Green's function. The Green's function equation for fermionic operator is
 
 \be
 M_{f}(r)G_{f}(r,r')\=\delta(r,r')\,\,I_{2k+2}\,.
 \ee
Now, following our assumption $(6)$, the fermionic Green's function can be obtained from the bosonic Green's function i.e. for $r<r'$, the fermionic Green's function is

 \begin{equation}\label{Green'sfn.fermion1}
 G^{<}_f(r,r')\equiv E(r)\hat G^{<}(r,r')E^{\dagger}(r') \= E(r) \begin{pmatrix} 
\frac{1}{\gamma_1} \wt G_1(r,r')& 0 &0\\
  0& \frac{1}{\gamma_2}\wt G_2(r,r')&0\\
  0&0& 0
 \end{pmatrix}  E^{\dagger}(r')\,,
 \end{equation}
 and for $r > r'$
 \begin{equation}\label{Green'sfn.fermion2}
 G^{>}_f(r,r')\equiv E(r)\hat G^{>}(r,r')E^{\dagger}(r') \= E(r)  \begin{pmatrix} 
\frac{1}{\gamma_1} \wt G'_1(r,r')& 0 &0\\
  0& \frac{1}{\gamma_2}\wt G'_2(r,r')&0\\
  0&0& 0
 \end{pmatrix} E^{\dagger}(r')\,.
 \end{equation}

Here, it is worth to mention couple of points. Firstly, the bosonic Green's function $\wt G_{1,2}$ and $\wt G'_{1,2}$ are such that the fermionic Green's function $G^{<}_f(r,r')$ and $G^{>}_f(r,r')$ satisfy the required boundary conditions as a function of both the argument $r$ and $r'$. Therefore, in general $\wt G_{1,2}$ and $\wt G'_{1,2}$ are different than $G_{1,2}$ and $G'_{1,2}$, respectively. In particular, it satisfies
\be
\hat M(r)\hat G(r,r')\=\delta(r,r')\begin{pmatrix}
 I_{k+1}& 0&0\\
 0&1 &0\\
  0&0&0
\end{pmatrix}\,. 
\ee
Now, when the boundary conditions are consistent with supersymmetry then one can see that given an admissible bosonic solution one can construct an admissible fermionic solution and vice versa.
Thus, for the supersymmetric boundary conditions we have $G_{1,2}(r,r')\=\wt G_{1,2}(r,r')$ and $G'_{1,2}(r,r')\=\wt G'_{1,2}(r,r')$. 
The argument for this is as follows:

Let us suppose that $s$, which is a $(k+2)$-vector, is a solution of the bosonic equation $M_b s =0$. Now, consider the~$(2k+2)$ dimensional vector $s_f = E\, \hat{s}$ where  $ \hat{s} = \begin{pmatrix}s\\{\bf 0}\end{pmatrix}$, where ${\bf 0}$ is a $k$ dimesnional zero. Then it follows that  $M_f s_f = (E^\dagger)^{-1} \hat M \hat{s}=0$. So for every bosonic solution $s^i$ we have the corresponding fermionic solution $s_f^i= E \hat{s}^i$. Of course, it goes other way also: for every fermionic solution $s_f^i$ , $P E^{-1} s_f^i$, where $P$ is the projector that projects to the first $(k+2)$ components, will be a bosonic solution. By supersymmetric boundary condition, it is meant that for every acceptable bosonic solution the corresponding fermionic solution is also acceptable (and of course this implies the other way also). Now, let us consider $G_b^{>}(r,r')$. near $r=\infty$ this will be linear combinations of bosonic solutions that are acceptable at  $r=\infty$ . Then, $G_f^{>}(r,r')= E(r) G_b^{>}(r,r') E^{\dagger}(r')$. As a function of $r$ and $r'$ this will be linear combinations of fermionic solutions of $M_f$ and its conjugate, respectively. If boundary condition are supersymmetric then it is clear $G_f^{>}(r,r')$ will be the correct fermionic Green's function. If the boundary conditions are not supersymmetric then it must be that there is some bosonic solution, say $s_b^1$, which is not acceptable at $r=\infty$ but the corresponding fermionic solution $s_f^1$ is acceptable. So in $G_f^{>}(r,r')= E(r) G_b^{>}(r,r')E^{\dagger}(r')$ one will have to start with a ``bosonic Greens function'' which as a function of $r$ involves $s_b^1$ in order to get acceptable fermionic Green's function. However, the acceptable bosonic Green's function will be different as it should not involve $s_b^1$ as a function of~$r$.

Secondly, note that $E(r)$ and $E^{\dagger}(r')$ are differential operators. So, in the definition of $G_f$ above the $E(r)$ appearing on the left is a differential operator that acts on the argument~$r$ of $\wt G_{1,2}(r,r')$ and $\wt G'_{1,2}$,  while $E^{\dagger}(r')$ appearing on the right is a differential 
 operator in variable $r'$ and acts  on the argument $r'$ of $\wt G_{1,2}(r,r')$ and $\wt G'_{1,2}(r,r')$~(with~$\frac{d}{dr'} \rightarrow -\frac{d}{dr'} $). One can see this as follows:
 
 We start with the inhomogeneous equation
 \begin{equation}
  M_f \begin{pmatrix} 
 QX_0\\
  c\\X_1'
\end{pmatrix} \= \begin{pmatrix} 
 h_1\\
  h_2\\h_3
\end{pmatrix}\,. 
 \end{equation}
Then we want to show that the solution of the above equation is 
\begin{equation}
  \begin{pmatrix} 
 QX_0(r)\\
  c(r)\\X_1'(r)
\end{pmatrix} \=\int_{r'>r}dr' G^{<}_f(r,r') \begin{pmatrix} 
 h_1(r')\\
  h_2(r')\\h_3(r') \end{pmatrix} +\int_{r'<r}dr' G^{>}_f(r,r')\begin{pmatrix} 
 h_1(r')\\
  h_2(r')\\h_3(r')
 \end{pmatrix}\,,
 \label{solution}
 \end{equation}
 with the functions~$G^{<}_f(r,r')$ and~$G^{>}_f(r,r')$ given in~\eqref{Green'sfn.fermion1} and~\eqref{Green'sfn.fermion2}, respectively.  
To prove this we first integrate $\frac{d}{dr'}$ appearing in  $E^{\dagger}$ in $G^{<}_f$ and $G^{>}_f$ by parts.
We get two contributions:
\ndt 1) The boundary term
\begin{eqnarray}
E_{1}(r)(\p_{r}\hat G^{>}_f(r,r')|_{r'=r-}-\p_{r}\hat G^{<}_f(r,r')|_{r'=r+})E^{\dagger}_{1}(r)\begin{pmatrix} 
 h_1(r)\\
  h_2(r)\\h_3(r) \end{pmatrix} &=& \begin{pmatrix} 
 0\\
  0\\\frac{1}{\gamma_1} D^{-1} C_1 (A_1^{b(2)})^{-1}C_1^{\dagger} D^{-1}h_3(r) \end{pmatrix}\,, \nonumber\\&=&
  \begin{pmatrix} 
 0\\
  0\\D^{-1}h_3(r)\end{pmatrix}\,,
  \label{bt}
 \end{eqnarray}
 where in the first equality we have used the discontinuity relation (\ref{disc}) and~$A_1^{b(2)}(r)$ is the matrix coefficient of the second order differential operator~$A_1^{b}(r)$. In the second equality we use the fact that the first order 
 derivative in $E(r)$ and $E(r)^{\dagger}$ appears only in the off-diagonal blocks involving $C(r)$ and $C(r)^{\dagger}$ where
 \begin{equation}
  C(r)\= C_1(r)\frac{d}{dr} + C_0(r),\quad \text{and}\quad C(r)^{\dagger}\= -\frac{d}{dr}C_{1}(r)^{\dagger} + C_{0}(r)^{\dagger}  \label{C1}\,.
 \end{equation}
The second equality in (\ref{bt}) can be argued as follows.  $C$ and $K$ are  $k\times (k+1)$ matrix and $1 \times 
(k+1)$ matrices. We can define a $k$ dimensional space $V_1$  and a one-dimensional 
space $V_2$ which satisfy the conditions :
\begin{equation}\label{Decomp.ofVectorV}
 K V_1\=0,\quad C_1 V_2 \=0\,.
 \end{equation}
The fact that $V_2$ is one- dimensional follows from the non-degeneracy of  coefficient of the second derivative 
term in $A_1$ namely $\gamma_1 A_1^{(2)}= -K^{\dagger}  A_{22}^{(2)} K+C_1^{\dagger} D^{-1} C_1$. Now we can 
choose a 
basis for $(k+1)$ dimensional space (represented as $(k+1)$ dimesnional row vector) such that the first $V_1$ 
occupies the first $k$ elements while $V_2$  the 
last element. Then the  $C_1 \=(c_1\,\bf{0})$ where $c_1$ is a non-degenerate $(k\times k)$ matrix and 
$\bf{0}$ is the $k$ dimensional null vector. Furthermore $K$ is a $(k+1)$ dimensional row vector with the first 
$k$ elements being zero. 
It follows that 
$ \gamma_1 A_1^{(2)}|_{V_2}\= -K^{\dagger}  A_{22}^{(2)} K$ and 
$\gamma_1  A_1^{(2)}|_{V_1}\=c_1^{\dagger} D^{-1} c_1$. The last equality implies that
$(\gamma_1  A_1^{(2)})^{-1}|_{V_1}\=c_1^{-1} D (c_1^{\dagger})^{-1}$.
Thus in this basis we have:
\begin{eqnarray}
\frac{1}{\gamma_1} D^{-1} C_1 (A_1^{(2)})^{-1}C_1^{\dagger} D^{-1}
&=&\frac{1}{\gamma_1} D^{-1} c_1 (A_1^{(2)}|_{V_1})^{-1}c_1^{\dagger} D^{-1}\,,\nonumber\\
&=& D^{-1}\,,
\label{btproof}
\end{eqnarray}
which proves (\ref{bt}). Applying $M_f$ on (\ref{bt}) gives:
\begin{equation}
\begin{pmatrix}
C^{\dagger} D^{-1}h3\\
  0\\h_3(r)\end{pmatrix}  \label{bt1}\,.
\end{equation}

\ndt 2) the bulk term 

This is the same as (\ref{solution}) except that $E^{\dagger}$ appeaing $G_f$ and $G'_f$ act now to the right ie. on the source.
The bulk term can be rewritten as
\begin{equation}
E(r)\Big(\int_{r'>r}dr'\, \hat G^{<}(r,r') +\int_{r'<r}dr'\, \hat G^{>}(r,r')\Big) E^{\dagger}(r')\begin{pmatrix}
 h_1(r')\\
  h_2(r')\\h_3(r')\end{pmatrix}\,.
  \label{bulk.1}
\end{equation}
This is so because the boundary term that appears in pulling  the differential operator $E$ outside the integral
vanishes due the discontinuity relation (\ref{disc}).

Now, let us apply $M_f(r) = (E^{\dagger}(r))^{-1}\hat{M}(r) E(r)^{-1}$ on the bulk term (\ref{bulk.1}). First of all $ E(r)^{-1}$ 
removes $E(r)$ in (\ref{bulk.1}). The action of the operator $\hat{M}$ on $\hat G^{<}(r,r') $ and $\hat G^{>}(r,r') $
vanishes since $r\neq r'$.
So, the only possible contribution can come when one of the derivatives $\frac{d}{dr}$
in $\hat{M}$ acts on the limits of the integrations. Using the discontinuity relations (\ref{disc}) one can show that
this results in
\begin{equation}
 (E^{\dagger})^{-1}\begin{pmatrix}
 1& 0&0\\
 0&1 &0\\
  0&0&0
\end{pmatrix} E^{\dagger}(r)\begin{pmatrix}
 h_1(r)\\
  h_2(r)\\h_3(r)\end{pmatrix}  \=\begin{pmatrix}
 h_1(r)-C^{\dagger}D^{-1} h3(r)\\
  h_2(r)\\0\end{pmatrix}\,,
  \label{bulk}
\end{equation}
where we have used the explicit form of $(E^{\dagger})^{-1}$
\begin{equation}
 (E^{\dagger})^{-1}\=\begin{pmatrix}
 1& -f^{-1,\dagger} K^{\dagger}&C^{\dagger}D^{-1}\\
 0&f^{-1,\dagger} &0\\
  0&0&1\end{pmatrix}\,.
\end{equation}
Adding the two contributions (\ref{bt1}) and (\ref{bulk}), one finds that $M_f$ acting on the proposed solution 
(\ref{solution}) indeed reproduces the source.
 
\subsection{Variation of one loop determinant}\label{Sec.Var.OneLoop}

Now one can compute the variation of the one loop determinant\eqref{greendet.1} with respect to $\alpha$
\begin{equation}\label{Var.OneLoop}
\delta_{\a} \ln Z_{\rm 1-loop} (\a)\=\frac{1}{2}\int^{\infty}_{0} dr \lim_{r'\rightarrow r+}{\text {tr}}\Big( \delta_{\a} M_f (r)G^{<}_f(r,r') - \delta_{\a} M_b(r) G^{<}_b(r,r')\Big)\,.
\end{equation}
Here ``${\text{tr}}$'' is just the matrix trace and $\delta_{\a}\equiv \frac{\delta}{\delta\a}$. In the above, we have taken the limit $r'\rightarrow r_+$. Had we taken the limit $r'\rightarrow r_-$, the fermionic and bosonic Greens functions will be 
replaced by $G^{>}_f$ and $G^{>}_b$, respectively but we will see later that the final result does not change. The fermionic part 
of the variation after using the form of $M_f$ and $G_f$ and some algebra, is
\bea\label{TraceVar.Fermion}
{\text {tr}}( \delta_{\a} M_f(r) G^{<}_f(r,r'))&=&\text{ tr}\frac{1}{\gamma_1}\Big(\delta_{\a} A_{11}(r)\wt G_1(r,r') - \delta_{\a}(K^{\dagger} A_{22}(r) K) \wt G_1(r,r')\nn\\&&- \delta_{\a} B(r)D^{-1}(r) C(r) \wt G_1(r,r')
 -\delta_{\a} C(r) \wt G_1(r,r') C^{\dagger}(r')D^{-1}(r')\nn\\
 &&+\delta_{\a} D(r)D^{-1}(r)C(r)\wt G_{1}(r,r')C^{\dagger}(r')D^{-1}(r')\Big)\nonumber\\
 &&+ \text{tr}\frac{1}{\gamma_2}\delta_{\a}( A_{22}(r))f(r) \wt G_2(r,r')f^{\dagger}(r')\,, \nn\\
 &=&\text{tr}\Big(\delta_{\a} A_1^{b}(r) \wt G_1(r,r')+f^{-1,\dagger}(r)\delta_{\a} A^{b}_{2}(r)\wt G_2(r,r')f^{\dagger}(r')\Big)\nn\\
&& +\text{tr}\frac{1}{\g_{1}}\Big(B(r)\delta_{\a}(D^{-1}(r)C(r))\wt G_{1}(r,r')-\delta_{\a} C(r) \wt G_1(r,r') C^{\dagger}(r')D^{-1}(r')\nn\\
 &&+\delta_{\a} D(r)D^{-1}(r)C(r)\wt G_{1}(r,r')C^{\dagger}(r')D^{-1}(r')\Big)\,.
\eea
In the above we have used the relations~\eqref{RelationAij} and also the fact that $G_{1}(r,r')$ and $G_{2}(r,r')$ are Green's function for the kinetic operators $A^{b}_{1}(r)$ and $A^{b}_{2}(r)$, respectively. 
Thus, the fermionic contributions to the variation~\eqref{Var.OneLoop} is
\bea
\lim_{r'\rightarrow r+}{\text {tr}}( \delta_{\a} M_f(r) G_f(r,r'))&=&\text{tr}\Big(\delta_{\a} A_2^b(r) \wt G_2(r,r)+\delta_{\a} A_1^{b}(r) \wt G_1(r,r)\Big)\nn\\
&& +{\text{tr}}\frac{1}{\g_{1}}\Bigl(C^{\dagger}(r)\delta_{\a}(D^{-1}(r) C(r)) \wt G_1(r,r)\nn\\
&& -\delta_{\a}(D^{-1}(r) C(r)) \wt G_1(r,r) C^{\dagger}(r) \Big)\,.
\eea

Now, we see that the first two terms in the above equation cancel the bosonic 
variation if and only if the boundary conditions are consistent with supersymmetry i.e. when the fermionic Green's function is contructed from the bosonic Green's function~\eqref{bosonicGreenfn.1} and~\eqref{bosonicGreenfn.2}. In this case, we are finally left with
\bea\label{Var.OneLoop.2}
 \delta_{\a} \ln Z_{\rm 1-loop} (\a)&=&
\frac{1}{2}\int^{\infty}_{0} dr\, {\rm tr}\frac{1}{\g_{1}}\Big(C^{\dagger}\delta_{\a}(D^{-1} C) G_1(r,r)
 -\delta_{\a}(D^{-1} C) G_1(r,r) C^{\dagger}) \Big)\,,\nn\\
\eea
where the differential operators $C$ and $C^{\dagger}$ appearing on the left and right of $G_1$ act on respectively 
the first and second arguments of $G_1$.
We can now move the operator $C^{\dagger}$ appearing on the right of $G_1$ in the 
second term of~\eqref{Var.OneLoop.2} to the left of $G_1$ by using cyclicity of matrix trace as well as an integration by part.  
This results in a bulk term which cancels with the first term and a boundary term. Thus, the variation of the one loop determinant becomes
\begin{equation}\label{boundaryterm.1}
\boxed{ 
\delta_{\a} \ln Z_{\rm 1-loop} (\a)\=-\frac{1}{2}{\rm tr}\frac{1}{\gamma_1}(C_1^{\dagger}(r)\delta_{\a}(D^{-1}(r) C(r)) G_1(r,r) \Big|^{r \rightarrow \infty}_{r\rightarrow 0}\,.}
\end{equation}
Note that the operator $C(r)$ acts only on the first argument of the Green's function. Thus, we find that if the fermionic Green's function are related to the bosonic Green's function as in~\eqref{Green'sfn.fermion1} and~\eqref{Green'sfn.fermion2}, the variation of the one loop determinant receives contribution only from the boundary. Moreover, to evaluate the boundary term, we just need to know the bosonic Green's function $G_{1}(r,r')$. This is one of the most important results of our paper.

Now, if it turns out that the $C$ is independent of $\alpha$ (as we will see in the examples of AdS$_{2}\times$S$^{1}$) or its $\a$ dependence is subleading near each boundary (we have also observed this in other examples \cite{dggn}), then from~\eqref{RelationAij}
we see that $D^{-1} \= \gamma_1 D_0^{-1}$,
where $D_0$ is independent of $\alpha$ (at least near each boundary). Using the relation $\frac{1}{\gamma_1}=Q^2$ we then conclude that
\begin{equation}\label{boundaryterm.2}
\delta_{\alpha} \ln Z_{\rm 1-loop} (\a)\=\frac{1}{2}(\delta_{\alpha} \ln Q^2){\rm Tr}(C_1^{\dagger}D_0^{-1} C) G_1(r,r)\Big |^{r \rightarrow \infty}_{r\rightarrow 0}\,.
\end{equation}
The above result \eqref{boundaryterm.2} we arrived at by taking the limit $r'\rightarrow r^+$. If we had taken the other  limit $r'\rightarrow r^-$, we would end up with the same expression as above but with $G_1$ replaced by $G'_1$. The difference between the variations will be
\be
\frac{1}{2}(\delta_{\alpha} \ln Q^2){\rm Tr}(C_1^{\dagger}D_0^{-1} C) (G_1(r,r)-G'_{1}(r,r))\Big |^{r \rightarrow \infty}_{r\rightarrow 0}\,.
\ee
Using the discontinuity relation of the Green's function, we find that the above difference becomes
\be
\frac{1}{2}(\delta_{\alpha} \ln Q^2){\rm Tr}(C_1^{\dagger}D_0^{-1} C_{1})  \frac{1}{A^{(2)}_{1}(r)}\Big |^{r \rightarrow \infty}_{r\rightarrow 0}=\frac{1}{2}(\delta_{\alpha} \ln Q^2){\rm Tr}_{V_{1}}I_{k}\Big |^{r \rightarrow \infty}_{r\rightarrow 0}=0\,.
\ee
In the above $I_{k}$ is a $k\times k$ identity matrix. Thus, it is reassuring that the result does not depend on the way one takes the limit $r' \rightarrow r$. 

It will be interesting to investigate the cases where the $\a$ dependence in $C$ is not subleading and its implications on the Green's function method presented above.

\subsection{Calculation of the boundary terms} \label{calcbterm}
Now, we will evaluate the boundary terms~\eqref{boundaryterm.1}. The boundary term is given in terms of the Green's function of the differential operator $A_1^b $ which is a $(k+1) \times (k+1)$ matrix second order differential operator. 
We have stated earlier, as a part of our assumptions~\eqref{RelationAij}, that the~$A_1^b $ can be 
expressed in terms of the fermionic operator as $A_{11}-K^{\dagger}A_{22}K-BD^{-1}C$. This is one of the consequences of supersymmetry. Furthermore, the second order derivative term in~$A_1^b $ comes from $K^{\dagger}A_{22} K$ and $BD^{-1}C $. While the former has rank 1 the latter has rank $k$. In order to simplify the computations, we can decompose the $(k+1)$ dimensional space in terms of a $k$ dimensional space $V_1$  and a one-dimensional space $V_2$ as in~\eqref{Decomp.ofVectorV}.
This means that second derivative part of $K^{\dagger}  A_{22} K$ in $A_1^b $ acts only on $V_2$ and 
that of $B D^{-1} C $, namely $B_1 D^{-1} C_1 $ acts only on $V_1$. Of course the first order 
derivative and non-derivative pieces contained in $A_{11}$ and  $B D^{-1} C $ will in general act on both 
$V_1$ and $V_2$ and therefore, the operator $A_1^b $ will mix these two spaces through lower order derivative terms.
To evaluate the boundary term~\eqref{boundaryterm.1} we will make the following assumptions. 
\begin{enumerate}
\item The leading behaviour of the solutions of  $A_1^b $ near the boundaries, i.e. near $r=0$ and $r=\infty$, is
determined by $K^{\dagger}  A_{22} K$ on $V_2$ and by $B D^{-1} C $ restricted to $V_1$. This means that 
the first order derivative and non-derivative pieces contained in $A_{11}$, $A_{22}$ and  $B D^{-1} C $ that mix $V_1$ and $V_2$ only contribute to subleading orders. We have checked in all the examples we have studied, assumption holds. In fact, our preliminary calculations also  indicate that 
the assumption follows  from  the general  positive definite localising action of the form
$S \sim \Psi (Q \Psi)^\dagger $. 
Therefore, to compute the boundary term~\eqref{boundaryterm.1} or~\eqref{boundaryterm.2}, we only need to study the action of $B D^{-1} C $ and the Green's function, $G_{1}(r,r')$, restricted to the vector space $V_{1}$. That is, the leading contribution to the boundary term comes from the space of solutions of $B D^{-1} C $ (now viewed as $(k \times k)$ matrix operator) on $V_1$.
\item The Greens fn for $A_1^b$ exists. This, taking into account assumption (1), implies that of the $2k$ 
solutions of $B D^{-1} C $ on $V_1$  near the boundary at least $k$ solutions satisfy the boundary conditions. 
Similarly, it implies that of the 2 solutions of $K^{\dagger}  A_{22} K$ on $V_2$ at least one solution satisfies the boundary condition.
\item $A_1^b$ has  no zero modes \footnote{ $A_2^b$ can have zero mode corresponding to the 
variation of the 
saddle point, which happens only for modes that are constant along space orthogonal to $AdS_2$ and  for which we 
already have collective coordinate integration. This case will be discussed separately.}.  
This means that  there are precisely $k$ 
solutions to $B D^{-1} C $ on $V_1$ and 1 solution of   $K^{\dagger}  A_{22} K$ on $V_2$  that are 
allowed near each of the boundaries and that none of the allowed $k$ 
solutions near one boundary, when analytically  continued to the other boundary satisfies the 
corresponding boundary condition.  
\end{enumerate}
In the following, by a slight abuse of notation, we will denote by $C$ and $B(= C^{\dagger})$ their restrictions to $V_1$ i.e. they will be represented (by a suitable change of basis) as $(k \times k)$ matrix operators, unless stated explicitly otherwise. Similarly, we will denote the Green's function of $A^{b}_{1}$ restricted to $V_{1}$ by $G_{1}(r,r')$ for $r<r'$ and $G'_{1}(r,r')$ for $r>r'$ and both will be a $k\times k$ matrix.

Now, the assumption (1) could have been relaxed. Of course even if this assumption is not valid in some cases, one can carry out the boundary analysis of the Green's functions and compute the boundary term in the $\alpha$ variation above in each case separately, but this assumption will allow us to obtain a general formula for the boundary term and relate it to the index of the differential operator  $C$. 

We begin with $2k$ linearly independent solutions of $A^{b}_{1}$ (now viewed as $(k \times k)$ matrix operator)
on $V_1$. 
Let us denote by $S$ a $(k \times 2k)$ matrix where the $2k$ columns label the $2k$ different solutions and let $\eta$ be a diagonal  $(2k \times 2k)$ 
matrix with 
entries $-1$ for the allowed solutions and $+1$ for the ones that are 
not allowed. From the assumptions (2) and (3), there are $k$ solutions each with $+1$ and $-1$ eigenvalues of $\eta$.  Thus, $\frac{1}{2}(1-\eta)$ and $\frac{1}{2}(1+\eta)$ are projections operator which will project the solutions matrix $S$ into  the acceptable and non-acceptable solutions near each boundary. Furthermore, the leading behaviour of the solution $S$ agrees with the leading behaviour of the solution of $BD^{-1}C$ restricted on $V_{1}$.
We define the Green's function to be
\begin{eqnarray}
 G_{1}(r,r') &=& \frac{1}{2}S(r)(1-\eta)X(r')\,, ~~~~~~ {\rm for} ~~ r < r' \,,\nonumber \\
 G'_{1}(r,r') &=& (\frac{1}{2}S(r)(1+\eta)+\cdots)X(r')\,, ~~~~~~ {\rm for}~~  r > r'\,.
 \label{green}
\end{eqnarray}
Here $X(r')$ is an unknown $(2k \times k)$ matrix such that $(1-\eta)X(r') $ is admissible at  the other boundary i.e. at $r\rightarrow\infty$ and  $(1+\eta)X(r') $ satisfy the allowed boundary condition at the first boundary i.e. at $r=0$.
The dots in the second equation above denote  combinations of allowed solutions i.e. 
of the form   $Y_1 \frac{1}{2}S(r)(1-\eta)Y_2 (1+\eta)$ where $Y_1$and $Y_2$  are some constant  
(i.e. independent of $r$) matrices. 
$Y_1$ and $Y_2$ are determined by requiring that the combination $ (\frac{1}{2}S(r)+Y_1 \frac{1}{2}S(r)
(1-\eta)Y_2) (1+\eta)$ are the analytic continuation of allowed solutions near the other boundary.  However these 
dotted terms will be subleading and therefore not be relevant for us and we will drop them in the following. 
What is important, however, is that the $k$ linearly
independent solutions that are admissible at the other boundary, let say at $r=0$, when analytically continued to $r=\infty$, span the $k$ dimensional space  
$\frac{1}{2}S(r)(1+\eta)$ (of inadmissible solutions) near the first boundary, as is implied by 
the assumption (3) of the non-existence of zero modes for $A_1^b$.  

Next, we determine $X(r)$. When $r \neq r'$ both $G$ and $\hat{G}$ are annihilated by $A_1^b$. The continuity/ discontinuity relations for the Greens function near $r=r'$ are:
\begin{eqnarray}\label{cont/discont}
&&G_{1}'(r,r)- G_{1}(r,r)\=0\,,\nonumber\\
&&\lim_{\epsilon\rightarrow 0}B_1 D_0^{-1} C_1\partial_r(G_{1}'(r,r')|_{r'=r-\epsilon}- G_{1}(r,r')|_{r'=r+\epsilon})\= {\bf 1}\,.
\end{eqnarray}
Here $\bf 1$ is a $k\times k$ identity matrix. Note that in the second line we have used the fact that second order differential operator $A_{22}$ in $A^{b}_{1}$ does not play a role on the solution in vector space $V_{1}$. Using the continuity equation, the discontinuity equation can also be written as
\be
\lim_{\epsilon\rightarrow 0}B_1 D_0^{-1} C(G_{1}'(r,r')|_{r'=r-\epsilon}- G_{1}(r,r')|_{r'=r+\epsilon})= {\bf 1}\,.
\ee
Using the expressions for the Green's function given in~\eqref{green}, we write the two equations in~\eqref{cont/discont} as a matrix equation for $X(r)$
\begin{equation}
  W(r)\eta X(r) \=  \begin{pmatrix}0 \\ {\bf 1}\end{pmatrix}\,,
  \label{eqW}
\end{equation}
where
\begin{eqnarray}
W \=\begin{pmatrix}
  S(r)\\
   B_1 D_0^{-1} C S(r)
 \end{pmatrix} =\begin{pmatrix}
 s(r)& \tilde{s}(r)\\
 B_1 D_0^{-1} C s(r)& B_1 D_0^{-1} C \tilde{s}(r)
\end{pmatrix}\,.
\end{eqnarray}
In the above we have split $k\times 2k$ matrix $S$ as $S=(s(r)\,\,\, \wt s(r))$, where $s(r)=\{s^{i}(r)\}$ and $\wt s(r)=\{\wt s^{i}(r)\}$, for $i=1,...,k$ are solutions of $A^{b}_{1}$. Thus we obtain
\be\label{eqX}
X(r)\=\eta W^{-1} \begin{pmatrix}0 \\ {\bf 1}\end{pmatrix}\,.
\ee
Note that the inverse of $W$ exist because the determinant of $W$ is
determinant of $B_1 D_0^{-1} C_1$ times the Wronskian and hence non-zero because of our assumptions. Since to evaluate the boundary term~\eqref{boundaryterm.2}, we just need to know the asymptotic form of the Green's function, we therefore, only require the asymptotic form of $X(r)$ at each boundary.  
To begin with we first consider the analysis near the boundary i.e. $r=0$. Without loss of generality, we can assume that the set of solutions $\{s^{i}(r)\}$, for $i=1,..,k$ belong to the kernel of $C$ near $r=0$. In this case near $r=0$, we have
\begin{eqnarray}
 \lim_{r\rightarrow 0}W^{-1}\= \begin{pmatrix} s^{-1} &  - s^{-1}\tilde{s}( B_1 D_0^{-1} C \tilde{s}(r))^{-1}
 \\0 &( B_1 D_0^{-1} C \tilde{s}(r))^{-1} \end{pmatrix}\,,
\end{eqnarray}
where we have used the fact that $ B_1 D_0^{-1} C s(r)=0$.  
In this case near the boundary $r=0$, the solution~\eqref{eqX} becomes
\begin{eqnarray}
 X(r) &=& \eta_{0} \begin{pmatrix}
 s^{-1}(r)& -s^{-1}(r)\tilde{s}(r)( B_1 D_0^{-1} C \tilde{s}(r))^{-1}\\
 0&( B_1 D_0^{-1} C \tilde{s}(r))^{-1}\end{pmatrix}\begin{pmatrix}0\\{\bf 1}
\end{pmatrix}\,,\nonumber\\
 &=&\eta_{0}\begin{pmatrix} -s^{-1}(r)\tilde{s}(r)( B_1 D_0^{-1} C \tilde{s}(r))^{-1} \\( B_1 D_0^{-1} C \tilde{s}(r))^{-1}\end{pmatrix}\,,
 \label{eqX}
\end{eqnarray}
where with this ordering of the solutions in $S(r)$, so that first $k$ column belongs to the Kernel of $C$ near $r=0$, the corresponding projector is $\eta_{0}$. Using the above equation we can obtain the Green's function for $r<r'$ near the boundary $r=r'=0$ as
\begin{equation}
 G_{1}(r,r')\=-\frac{1}{2} \begin{pmatrix} s(r)& \tilde{s}(r)\end{pmatrix}(1-\eta_{0})
 \begin{pmatrix} -s^{-1}(r')\tilde{s}(r')( B_1 D_0^{-1} C \tilde{s}(r'))^{-1} \\( B_1 D_0^{-1} C \tilde{s}(r'))^{-1}\end{pmatrix}
 \label{eqX}\,.
\end{equation}

Now, we can compute the boundary term at $r=0$ by using (\ref{boundaryterm.2}) and the expression for $G(r,r')$ from (\ref{green}) and 
(\ref{eqX}) and the result is
\begin{eqnarray}
\lim_{r\rightarrow 0}\text{Tr}(B_1 D_0^{-1} C G_{1}(r,r')|_{r'\rightarrow r} ) &=& 
-\frac{1}{2}\text{Tr}\Big[\begin{pmatrix}0&
B_1 D_0^{-1} C \tilde{s}(r)\end{pmatrix}(1-\eta)\begin{pmatrix} -s^{-1}(r)\tilde{s}(r) ( B_1 D_0^{-1} C \tilde{s}(r))^{-1}
\\( B_1 D_0^{-1} C \tilde{s}(r))^{-1}\end{pmatrix}\Big]\,,\nonumber\\ 
&=&-\frac{1}{2}\text{Tr}\Big[(1-\eta)\begin{pmatrix}0&-s^{-1}(r)\tilde{s}(r)  
\\0&\bf 1\end{pmatrix}\Big]\,,\nonumber\\
&=&-( k-\ell)\,.
\end{eqnarray}
where $\ell$ is the number of admissible solutions at $r=0$ that are in the Kernel of $C$.

We can repeat the same analysis at $r={\infty}$. The difference now is that for $r > r'$ the Green's function must involve solutions that are admissible near $r=\infty$. Let the corresponding projector be $\frac{1}{2}(1-\eta_{\infty})$. Then we have the Green's function as in \eqref{green} with $\eta\rightarrow -\eta_{\infty}$. We can repeat the above analysis except that we assume that our set of solutions to  $S(r)=(s'(r)\,\,\,\wt s'(r))$ such that the first $k$ column belongs to the Kernel of $C$ near $r=\infty$. Following the same steps as above, we get the contribution to the boundary term near $r=\infty$
 \begin{equation}
  -\frac{1}{2}\text{Tr}\Big[(1+\eta_{\infty})\begin{pmatrix}
0&-s^{-1}(r)\tilde{s}(r)  
\\0&\bf 1\end{pmatrix}\Big]\nonumber\\
\=-\ell'\,,
 \end{equation}
 where $\ell'$ is the number of admissible solutions in the set $\{s'(r)\}$ at $r=\infty$ that are in the Kernel of $C$. 
 Note in the above we have used the fact that the first $k\times k$ block of $\frac{1}{2}(1+\eta_{\infty})$ has $\ell'$ zeros.                                                                       

Taking the difference between the contribution at $r=\infty$ and at $r=0$ one ends up with the simple result
\begin{equation}\label{IndexC.1}
{\rm BT}\equiv {\rm Tr}(C_1^{\dagger}D_0^{-1} C) G_1(r,r)\Big |^{r \rightarrow \infty}_{r\rightarrow 0}\= (k-\ell-\ell')\,.
\end{equation}
Finally  combining (\ref{IndexC.1}) and (\ref{boundaryterm.2})  we obtain our main result 
\begin{equation}\label{main_result}
\boxed{ 
\delta_{\alpha} \ln Z_{\rm 1-loop} (\a)\=\frac{1}{2}(\delta_{\alpha} \ln Q^2) (k-\ell-\ell')\,. }
\end{equation}

Here we again recall the various integers that occur in this expression.
\begin{enumerate}
\item
$k+1$ is the integer that defines dimension of the bosonic space $X_0$. 
\item
$\ell, \ell'$ are the number of admissible solutions of the first order equations $Cs(r) = 0$ at the origin
and at asymptotic infinity of AdS, respectively. 
\end{enumerate}
Note that the above result \eqref{main_result} is obtained for each Kaluza Klein mode. Therefore, to obtain the complete contribution to the variation of the one loop determinant we need to sum over KK modes labelled by $\vec n$
\be
\delta_{\alpha} \ln Z^{\text{total}}_{\rm 1-loop} (\a)\=\frac{1}{2}\sum_{\vec n}(\delta_{\alpha} \ln Q_{\vec n}^2) (k-\ell_{\vec n}-\ell_{\vec n}')\,.
\ee
 \subsection{Connection to index of the operator $C$}\label{IndexofC}
In this section we will show that the result of the boundary term~\eqref{IndexC.1} is an index of the first order differential operator $\mathcal C=C|_{V_{1}}$. To show this we start with the fact that the operator $A^{b}_{1}(r)\Big|_{V_{1}}$, whose one loop determinant we are interested in to compute, asymptotically approaches $\mathcal C^{\dagger}D^{-1}\mathcal C$. Therefore, a solution of the operator $A^{b}_{1}(r)\Big|_{V_{1}}s=0$ near each boundary either belong to the solution space $S(\mathcal C)$ of the operator $\mathcal C$ or the solutions space $S(\mathcal C^{\dagger})$ of $\mathcal C^{\dagger}$ which is a subset of the image of the operator $\mathcal C$ .
We start with the solution space $S(\mathcal C)$. A solution in $S(\mathcal C)$ has asymptotic behaviour $r^{\g}$ and $e^{\hat\g r}$ near $r\rightarrow 0$ and $r\rightarrow\infty$, respectively. Let this set be $\Gamma(\mathcal C)=\{\g_{1},....,\g_{k}\}$ and $\hat\Gamma(\mathcal C)=\{\hat\g_{1},....,\hat\g_{k}\}$. The rest $k$ solutions of $A^{b}_{1}(r)\Big|_{V_{1}}$ correspond to the set $S(\mathcal C^{\dagger})$, the space of solution of $\mathcal C^{\dagger}$, and the corresponding set of the asymptotic behaviour be $\Gamma(\mathcal C^{\dagger})=\{\g^{*}_{1},....,\g^{*}_{k}\}$ and $\hat\Gamma(\mathcal C^{\dagger})=\{\hat\g^{*}_{1},....,\hat\g^{*}_{k}\}$. Now given these sets near each boundary the differential operators $\mathcal C$ and $\mathcal C^{\dagger}$ can be diagonalised. Near $r\rightarrow 0$ differential operators $\mathcal C$ and $\mathcal C^{\dagger}$  can be brought to the form\footnote{Note that one can always put the operator $\mathcal C$ and $\mathcal C^{\dagger}$ of the form \eqref{Asympt.ofC.1} and \eqref{Asympt.ofC.2} without the non derivative term being diagonal. }
\be\label{Asympt.ofC.1}
\mathcal C=I_{k}\frac{d}{dr}+\frac{1}{r}\mathcal C^{\text{diag.}}_{0},\quad \mathcal C^{\dagger}=-I_{k}\frac{d}{dr}+\frac{1}{r}\mathcal C^{\dagger\text{diag.}}_{0}\,,
\ee
and near $r\rightarrow \infty$ differential operators $\mathcal C$ and $\mathcal C^{\dagger}$  can be brought to the form
\be\label{Asympt.ofC.2}
\mathcal C=I_{k}\frac{d}{dr}+\mathcal C^{\text{diag.}}_{\infty},\quad \mathcal C^{\dagger}=-I_{k}\frac{d}{dr}+\mathcal C^{\dagger\text{diag.}}_{\infty}\,.
\ee
Here $\mathcal C^{\text{diag.}}_{0} (\mathcal C^{\dagger\text{diag.}}_{0})$ and $\mathcal C^{\text{diag.}}_{\infty}(\mathcal C^{\dagger\text{diag.}}_{\infty})$ are constant $k\times k$ matrices with diagonal entries given by $\Gamma(\mathcal C)\, (\Gamma(\mathcal C^{\dagger}))$ and $\hat\Gamma(\mathcal C)\,(\hat\Gamma(\mathcal C^{\dagger}))$, respectively.  

 Next, we consider an operator $\mathcal C' (\mathcal C^{'\dagger})$ which is continuously connected to $\mathcal C (\mathcal C^{\dagger})$ and is defined globally for every value of $r$.  This operator has the form
 \be
 \mathcal C'=I_{k}\frac{d}{dr}+\mathcal C_{0}'(r),\quad \mathcal C'^{\dagger}=-I_{k}\frac{d}{dr}+\mathcal C'^{\dagger}_{0}(r)\,.
 \ee
The non derivative term $\mathcal C_{0}' (\mathcal C_{0}^{'\dagger})$ is such that the operator $\mathcal C' (\mathcal C^{'\dagger})$ is a $k\times k$ diagonal first order differential operator for every value of $r$ and near the boundary it approaches the asymptotic form \eqref{Asympt.ofC.1} and \eqref{Asympt.ofC.2} of the differential operator $\mathcal C (\mathcal C^{\dagger})$. Thus, $\mathcal C' (\mathcal C'^{\dagger})$ is an interpolating operator between the asymptotic \eqref{Asympt.ofC.1} and \eqref{Asympt.ofC.2}. Since the operator $\mathcal C'$ is continuously connected to the operator $\mathcal C$, we expect that the index of $\mathcal C'$ to be same as that of the  operator $\mathcal C$.
 
Now we will compute the index of the operator $\mathcal C'$. Let $S(\mathcal C')$ be the space of solutions of matrix differential operator $\mathcal C'$. Since $\mathcal C'$ is a $k\times k$ first order matrix differential operator, we expect the dimension for the space of solutions to be $\text{dim}\,S(\mathcal C')=k$. We consider two spaces, $S_{1}(\mathcal C')\subset S(\mathcal C')$ and $S_{2}(\mathcal C')\subset S(\mathcal C')$, where $S_{1}(\mathcal C')$ are the set of solutions which are smooth near $r=0$ and $S_{2}(\mathcal C')$ are the set of solutions which are admissible near $r=\infty$. Since, operators $\mathcal C$ and $\mathcal C'$ have the same asymptotic, therefore, they have the same dimension of the space of admissible solution. Thus, $\text{dim}\,S_{1}(\mathcal C')=\ell$ and $\text{dim}\,S_{2}(\mathcal C')=\ell'$.   Let the space of Kernel of $\mathcal C'$ is $\text{Ker}(\mathcal C')$ and its dimension is $s$. The space $\text{Ker}(\mathcal C')\subset S(\mathcal C')$ is the space of solutions which are smooth near $r=0$ as well as admissible near $r=\infty$. Clearly, $\text{Ker}(\mathcal C')=S_{1}(\mathcal C')\cap S_{2}(\mathcal C')$. Furthermore, we expect that there are solutions $\in S(\mathcal C')$ which are neither smooth near $r=0$ nor admissible near $r=\infty$. These solutions belong to the space $\hat S(\mathcal C')=S(\mathcal C')/ S_{1}(\mathcal C')\cup S_{2}(\mathcal C')$ and the dimension of this space is
\be
\text{dim}\,\hat S(\mathcal C')=k-\ell-\ell'+s=k-\ell-\ell'+\text{dim}\,\text{Ker}(\mathcal C')\,.
\ee
Next, we argue that for every solution belonging to $ \hat S(\mathcal C')$, $\exists$ a solution belonging to $\text{Ker}(\mathcal C^{'\dagger})$. In particular, given a solution in $S(\mathcal C')$ which is neither smooth near $r=0$ nor admissible near $r=\infty$, the existence of Green's function of $A^{b}_{1}(r)$ requires that there exist a solution belonging to the Kernel of $\mathcal C^{\dagger}$ which is smooth near $r=0$ and admissible near $r=\infty$. Thus
\be
\text{dim}\,\text{Ker}(\mathcal C^{'\dagger})-\text{dim}\,\text{Ker}(\mathcal C')=k-\ell-\ell'.
\ee
The argument goes as follows: Let us consider a solution $s_{i}\in \hat S(\mathcal C')$ which has asymptotic determined by  $\g_{i}\in \Gamma(\mathcal C)$ and $\hat\g_{i}\in\hat\Gamma(\mathcal C)$ near $r=0$ and $r=\infty$, respectively. Both $\g_{i}$ and $\hat\g_{i}$ correspond to non admissible behaviour. Now we require that near each boundary the Green's function of $A^{b}_{1}(r)\Big|_{V_{1}}$ exists. Since, $A^{b}_{1}(r)\Big|_{V_{1}}$ asymptote to $\mathcal C^{\dagger}D^{-1}\mathcal C$, it implies that for every such $\g_{i}\in \Gamma(\mathcal C)$ at $r=0$ there is $\g^{*}_{i}\in\Gamma(\mathcal C^{\dagger})$ and for every such $\hat\g_{i}\in \hat\Gamma(\mathcal C)$ at $r=\infty$ these is $\g^{*}_{i}\in\hat\Gamma(\mathcal C^{\dagger})$, where $\g^{*}_{i}$ and $\hat\g^{*}_{i}$ give rise admissible asymptotic behaviour. Since $\mathcal C^{'\dagger}$ asymptote to $\mathcal C^{\dagger}$ near each boundary, this implies that there exist a solution $\bar s_{i}$ of $\mathcal C^{'\dagger}$ which has the asymptotic behaviour determined by $\g^{*}_{i}$ and $\hat\g^{*}_{i}$ and is acceptable at both ends. Thus it belongs to the kernel of $\mathcal C^{'\dagger}$. 
Furthermore, using the inner product $<v_{1},v_{2}>=\int dr\,v^{\dagger}_{1}v_{2} $, one sees that the space $\text{Ker}\,\mathcal C^{'\dagger}$ is isomorphic to the space $\text{Coker}\,\mathcal C'$. Thus 
\be
\text{ind}(\mathcal C')\equiv\text{dim}\,\text{Coker}(\mathcal C')-\text{dim}\,\text{Ker}(\mathcal C')=k-\ell-\ell'.
\ee
Since $\mathcal C'$ is continuously related to $\mathcal C$, 
\be
\text{ind}(\mathcal C)=k-\ell-\ell'.
\ee
Thus, the boundary term \eqref{IndexC.1} is the index of the operator $\mathcal C=C|_{V_{1}}$.

\section{Chern-Simons theory on $AdS_{2}\times S^{1}$: Greens function approach}\label{chern-section}
In this section, we revisit the analysis presented in \cite{David:2016onq}. In \cite{David:2016onq}, we computed the partition function of a non abelian bosonic Chern Simons theory on the metric background  
\be
ds^{2}=d\tau^{2}+L^{2}(dr^{2}+\sinh^{2}r\,d\theta^{2})\,,
\ee
where $L$ is some constant, using the supersymmetric localization. This is possible because of the following reason: The supersymmetric completion of a bosonic Chern-Simons action is
\be\label{SCSaction}
S_{\text{C.S.}}=\int d^3x\sqrt{g}\text{Tr}\left[i\varepsilon^{\mu\nu\rho}\left(a_\mu\p_\nu a_\rho-\frac{2i}{3}a_\mu a_\nu a_\rho\right)-\tilde\lambda\lambda+\frac{i}{2}H\sigma\right]\,.
\ee
Here $\varepsilon^{\mu\nu\rho}=\frac{1}{\sqrt{g}}\epsilon^{\mu\nu\rho},\quad \epsilon^{\tau\eta\theta}=1$. Also, in order to construct supersymmetric action, we have used the vector multiplet in $\mathcal N=2$ theory in Euclidean signature which contains an imaginary scalar $\sigma$, gauge field $a_\mu$, an auxiliary scalar field $H$ which is also imaginary and 2 component complex fermions $\lambda$ and $\tilde\lambda$. 
Now, we note that the fermions and scalars in the vector multiplet are purely auxiliary fields as they do not have kinetic terms and therefore, one can integrate them out. Thus the supersymmetric Chern-Simons theory is equivalent to a bosonic Chern-Simons theory. 

The analysis in \cite{David:2016onq} was based on index computation which relies on the boundary conditions being consistent with supersymmetry. These consist of normalizable boundary conditions on the gauge field and non normalizable boundary conditions on fermions following from supersymmetry transformations. 
We find that the one loop determinant evaluated using the index calculations is given as
\be
Z_{1-\text{loop}}(\a)=\prod_{\rho}\sqrt{\prod_{n\neq 0}(n-i\rho\cdot\a)\prod_{p\neq 0}(\frac{p}{L}-i\rho\cdot\a)}\,.
\ee

We will reproduce the above answer in the Green's function approach with normalizable boundary conditions on all fields, including fermions, and find that the above result holds true as long as $L^{2}>\frac{3}{4}$. It would be interesting to understand the significance of the rational number $\frac{3}{4}$.
\subsection{Q-exact deformation and gauge fixing}
Next, we deform the action \eqref{SCSaction} by a $Q$-exact term, $t\,QV_{{\rm loc} }$. 
We express the $QV_{{\rm loc} }$ in terms fermion bilinear $(\Psi,\Psi_\mu)$ instead of $(\lambda,\tilde\lambda)$ which are defined as 
\be
\Psi=\frac{i}{2}(\tilde\epsilon\lambda+\epsilon\tilde\lambda)\,,\quad \Psi_\mu=Q_{s}a_\mu=\frac{1}{2}(\epsilon\gamma_\mu\tilde\lambda+\tilde\epsilon\gamma_\mu\lambda)\,.
\ee
The fermion bi-linears are convenient for the evaluation of the index. 
The inverse of the above relations expresses $(\lambda,\tilde\lambda)$ in terms of  $\Psi,\Psi_\mu$ as
\bea
\lambda=\frac{1}{\tilde\epsilon\epsilon}\left[\gamma^\mu\epsilon\Psi_\mu-i\epsilon\Psi\right],\quad \tilde\lambda=\frac{1}{\epsilon\tilde\epsilon}\left[\gamma^\mu\tilde\epsilon\Psi_\mu-i\tilde\epsilon\Psi\right]\,.
\eea
The supersymmetry transformation of the bi-linears are
\bea
&&Q_{s}\Psi=\frac{1}{4}(\tilde\epsilon\epsilon)H-\frac{i}{2}\left(\tilde\epsilon\gamma^{\mu\nu}\epsilon\right) F_{\mu\nu}-\frac{1}{L}\sigma\,,\nn\\
&&Q_{s}\Psi_\mu=\mathcal L_K a_\mu+D_\mu\Lambda\,,
\eea
where $\Lambda=\wt\epsilon\epsilon\, \sigma-K^{\m}a_{\m}$.
One convenient choice of $V_{{\rm loc}}$ is given by
\be
V_{{\rm loc} }=\int d^3x\sqrt{g}\frac{1}{(\tilde\epsilon\epsilon)^2}\text{Tr}\left[\Psi^\mu (Q_{s}\Psi_\mu)^\dagger+\Psi (Q_{s}\Psi)^\dagger\right]\,.
\ee
The bosonic part of the $QV_{{\rm loc} }$ action is given by
\bea\label{bosonLagr}
Q_{s}V_{{\rm loc} \{\text{bosonic}\}}&=&\int d^3x\sqrt{g}\frac{1}{2(\tilde\epsilon\epsilon)^2}\text{Tr}\left[(Q_{s}\Psi^\mu)(Q_{s}\Psi_\mu)^\dagger+(Q_{s}\Psi)(Q_{s}\Psi)^\dagger\right]\,,\nn\\&=&\int d^3x\sqrt{g}\text{Tr}\Big[\frac{1}{4}F_{\mu\nu}F^{\mu\nu}-\frac{1}{2\cosh^2r}D_\mu(\cosh r\,\sigma)D^\mu(\cosh r\,\sigma)\nn\\
&&\qquad\qquad\qquad\qquad\qquad\qquad-\frac{1}{32}\left(H-\frac{4\sigma}{L\cosh r}\right)^2\Big]\,.\nn\\
\eea
For a gauge group $G$ with rank $r$, the minimum of the $Q_{s}V_{{\rm loc} \{\text{bosonic}\}}$ is parametrized by $r$ real parameters as
\be
a_\mu=0\,,\quad\sigma=\frac{i\alpha}{\cosh r}\,,\quad H=\frac{4i\alpha}{L\cosh^2r}\,.
\ee
Here $\alpha$ is a real constant matrix valued in Lie algebra of the gauge group. Furthermore, on this localization background the gauge transformation parameter in supersymmetry algebra reduces to a constant, $\Lambda^{(0)}=i\alpha$.\\
Next, we need to introduce the gauge fixing Lagrangian. In our case it turns out that the Green's function analysis becomes simpler for the gauge fixing Lagrangian
\be\label{gfLagrangian}
\mathcal L_{g.f.}=\text{Tr}\,Q\left[i(\tilde c\,\cosh^{2}r+2[\a,c])\nabla_\mu\left(\frac{1}{\cosh^2r}a^\mu\right)+\xi \tilde c b\right]\,,
\ee
where $Q=Q_{s}+Q_{B}$ and $Q_{B}$ is the BRST transformation. Below we will define the action of the supersymmetry transformations and BRST transformations on all the fields. 

Note that the above gauge fixing Lagrangian is different than the one used in \cite{David:2016onq}. As we will see below, the above choice of the gauge fixing Lagrangian decouples the equations of motion for the fluctuations of the scalar field $\sigma$ with the gauge field fluctuations. 

The complete action including the gauge fixing Lagrangian is invariant under BRST transformations on the fields which are given by
\bea
&&Q_Ba_\mu=D_\mu c,\quad Q_B\tilde c=b,\quad Q_B c=\frac{i}{2}\{c,c\},\quad Q_B\tilde\lambda=i\{c,\tilde\lambda\},\nn\\&&Q_B\lambda=i\{c,\lambda\},\quad Q_B\hat\sigma=i[c,\hat\sigma],\quad Q_B\hat H=i[c,\hat H],\quad Q_Bb=0\,.
\eea
Here $a_\mu,\,\hat\sigma$ and $\hat H$ are fluctuations away from localizing . \\
We also define the susy transformations for extra fields
\be\label{SusyTransf.ofGhost}
Q_{s}c=-\Lambda+\Lambda^{(0)},\quad Q_{s}b=\mathcal L_K\tilde c+i[\Lambda^{(0)},\tilde c],\quad Q_{s}\tilde c=0\,,
\ee
such that the combined transformations generated by $Q=Q_{s}+Q_B$ satisfy the algebra
\be\label{Qhatalgebra}
 Q^2=\mathcal L_K+\delta^{\text{gauge transf.}}_{\Lambda^{(0)}}\,.
\ee
To summarize, the complete transformations of fields under $\hat Q$ are given by
\bea\label{QhatTransf}
&& Qa_\mu=\Psi_\mu+D_\mu c,\qquad  Q\hat\sigma=Q_{s}\hat\sigma+i[c,\hat\sigma]\,,\nn\\
&& Q\Psi_\mu=\mathcal L_K a_\mu+D_\mu\Lambda+i\{c,\Psi_\mu\},\quad  Q\Psi=
\frac{1}{4}(\tilde\epsilon\epsilon)\hat H
-\frac{i}{2}\left(\tilde\epsilon\gamma^{\mu\nu}\epsilon\right) F_{\mu\nu}(a)-\frac{1}{L}\hat\sigma+i\{c,\Psi\}\,,\nn\\
&& Qc=-\Lambda+\Lambda^{(0)}+\frac{i}{2}\{c,c\},\qquad Q\tilde c=b\,.
\eea
At this point it is worth to mention a point which will be important in the later analysis. In our $\xi$-gauge, we see from the ghost Lagrangian involving fields $(\wt c, b)$ 
\be
{\cal L}_{\wt c,b}=\text{Tr}[ib\cosh^{2}r\nabla_{\m}(\frac{1}{\cosh^{2}r}a^{\m})-\xi b^{2}+i\wt c\cosh^{2}r\nabla_{\m}(\frac{1}{\cosh^{2}r}\p^{\m}c)+\xi\wt c({\cal L}_{K}\wt c+i[\Lambda^{(0)},\wt c])]\,,
\ee
that if we choose
\be\label{ZeroModeTildeC}
\wt c=\frac{\vec \m}{\cosh^{2}r},\quad b=\frac{\vec \m'}{\cosh^{2}r}\,,
\ee
where $\vec\m$ and $\vec\m'$ are gauge Lie algebra valued constant, then this mode decouples from the rest of the fields in the theory. The quadratic terms involving $b$ and $\wt c$ only gives a mass terms for this mode which is proportional to $\xi$. In fact, in $\xi=0$ limit these are zero modes. We will keep $\xi$ non zero for our convenience, however, we will subtract the contribution of this mode in the later calculation.

\subsection{Boundary conditions}
In this section we will discuss the boundary conditions on the fields present in the theory. This is essential when we define a quantum field theory on spaces with boundary. These boundary conditions set the value of the field at the boundary. In fact different boundary conditions define different quantum field theory. However, in the present case we are considering spaces which are of non compact type such as $AdS$. In this case the boundary conditions are much more reacher. $AdS$ space being an open space, one needs to impose conditions on the asymptotic behaviour of fields. Typically, these asymptotic fall off conditions on fields are motivated by preserving certain aspect of the theory such as preserving certain symmetry, normalizability and the ones motivated from the AdS/CFT correspondence. Here, we follow normalizability as the criteria on the fall off conditions i.e. we require that fluctuations of all the fields present in the theory on $AdS$ space should fall off asymptotically in a manner such that they are $L^{2}$-normalizable. Assuming this condition we find that for the bosonic fields in the vector multiplet, the fields should fall off asymptotically to satisfy
\bea\label{NormConditions.1}
e^{r/2}a_{t}\rightarrow 0,\quad e^{r/2}a_{r}\rightarrow 0,\quad e^{-r/2}a_{\theta}\rightarrow 0,\quad e^{r/2}\sigma\rightarrow0\,.
\eea
Here $a_{\m}$ and $\sigma$ are Lie algebra valued gauge field and scalar field, respectively. Similarly, requiring that the gaugino fields, $\lambda$ and $\wt\lambda$, are normalizable implies that 
\be\label{NormConditions.2}
\Psi_{t}\rightarrow 0,\quad \Psi_{r}\rightarrow 0,\quad e^{-r}\Psi_{\theta}\rightarrow 0,\quad \text{and}\quad \Psi\rightarrow 0\,.
\ee
Next, we want to define the boundary conditions on the ghost system. The ghost system consists of two grassmann odd scalar $c,\wt c$ and the Lagrange multiplier field $b$. The normalizable boundary condition on the Lagrange multiplier $b$ implies that the fluctuations should statisfy $e^{r/2}b\rightarrow 0$.
The boundary condition on the ghost field~$c$ is chosen to be the same as in~\cite{David:2016onq} i.e.
\be\label{NormConditions.3}
c\rightarrow f(\theta)+e^{-r/2}\wt f(\theta,\tau)+....\,.
\ee
This was motivated from the fact that $c$ is a gauge transformation parameter and we allow fluctuations of $c$ which does not change the boundary conditions on the gauge field. Once we have chosen the boundary conditions on the field~$c$, the boundary condition on the ghost~$\wt c$ is fixed by requiring that 
\be
\int d^{2}x\sqrt{g}\,\,\wt c\,c\,\,<\infty\,.
\ee
This requires that the field $\wt c$ should satisfy $e^{r}\,\wt c\rightarrow 0$, i.e. it falls faster than $e^{-r}$. Later on, we will see that these boundary condition on $c$ and $\wt c$ are essential in order to construct their Green's function.
\subsection{Equations of motions and the  Greens function} \label{eom-section}
As we explained earlier, the variation of the one loop determinant is given by the product of the variation of the differential operator and its Green's function. The differential operator appears at the quadratic order in the fluctuations in the~$QV$ action. The Green's function can be explicitly constructed out of the solutions of the equations of motions of the differential operator. However, in the supersymmetric case to evaluate the variation of the one loop determinant, we do not need the explicit form of these solutions rather only their asymptotic behaviour, which is a considerable simplifications. In this section, we will present these differential operator for both bosonic and fermionic fields and their Green's function. After this we will discuss the asymptotic behaviour of these differential operator which we will use to construct the asymptotic solutions. Furthermore, for the purposes of the presentation we will assume the Gauge group is $SU(2)$, but near the end we will generalize the result to any arbitrary compact group.

\subsubsection*{Equations of motions}
We begin with the bosonic fields. In the discussion below we will not care about the auxiliary field $H$, as its equation of motion is trivial and we assume that we have integrated it out in the path integral. The rest of the bosonic fields are the vector field $a_{\m}$ and the scalar field $\sigma$ which are elements in the Lie algebra of $SU(2)$. In the following discussion we will only consider the non-Cartan part of these fields. This is because the quadratic fluctuations containing the fields in the Cartan do not depend on $\alpha$ and thus, do not contribute to the variation in the one loop determinant. It is easy to see this in the bosonic action \eqref{bosonLagr} (and similarly for fermionic action).

We first expand the fields in terms of Fourier modes and write the Lagrangian in terms of the following Fourier modes
\bea
&&a^{1}_{t}\=\frac{1}{2}a^{+}_{t;n,p}(r)e^{i(nt+p\th)}+\frac{1}{2}a^{-}_{t;n,p}(r)e^{-i(nt+p\th)},\,\, a^{1}_{r}\=\frac{i}{2}a^{+}_{r;n,p}(r)e^{i(nt+p\th)}-\frac{i}{2}a^{-}_{r;n,p}(r)e^{-i(nt+p\th)}\nn\\
&&a^{1}_{\theta}\=\frac{1}{2}a^{+}_{\theta;n,p}(r)e^{i(nt+p\th)}+\frac{1}{2}a^{-}_{\theta;n,p}(r)e^{-i(nt+p\th)},\,\, a^{2}_{t}\=-\frac{i}{2}a^{+}_{t;n,p}(r)e^{i(nt+p\th)}+\frac{i}{2}a^{-}_{t;n,p}(r)e^{-i(nt+p\th)}\nn\\
&&a^{2}_{r}\=\frac{1}{2}a^{+}_{r;n,p}(r)e^{i(nt+p\th)}+\frac{1}{2}a^{-}_{r;n,p}(r)e^{-i(nt+p\th)},\,\, a^{2}_{\theta}\=-\frac{i}{2}a^{+}_{\theta;n,p}(r)e^{i(nt+p\th)}+\frac{i}{2}a^{-}_{\theta;n,p}(r)e^{-i(nt+p\th)}\nn\\
&&\sigma^{1}\=\frac{1}{2}\sigma^{+}_{n,p}e^{i(nt+p\th)}+\frac{1}{2}\sigma^{-}_{n,p}e^{-i(nt+p\th)},\,\, \sigma^{2}\=\frac{1}{2i}\sigma^{+}_{n,p}e^{i(nt+p\th)}-\frac{1}{2i}\sigma^{-}_{n,p}e^{-i(nt+p\th)}\,.
\eea
Here the labels on the fields are the usual labels of the Lie algebra $su(2)$.

The equations of motion for the vector field and scalar field are obtained by varying the action with respect to $a^{-}_{\m;n,p}$ and $\sigma^{-}_{n,p}$ and can be written as
\be
\mathcal M_{b}E_{b;n,p}^{+}(r)\equiv M_{2}\p^{2}_{r}E_{b;n,p}^{+}(r)+M_{1}\p_{r}E_{b;n,p}^{+}(r)+M_{0}E_{b;n,p}^{+}(r)\=0\,.
\ee
Here $M_{2,1,0}$ are $4\times 4$ matrices whiose elements are functions of coordinate $r$. The explicit form of these matrices are given in Appendix~\ref{appenB}. The column vector $E_{b;n,p}^{+}(r)$ is given as
\be
E_{b;n,p}^{+}(r)\=\begin{pmatrix}a^{+}_{t;n,p}(r)\\a^{+}_{r;n,p}(r)\\a^{+}_{\th;n,p}(r)\\\sigma^{+}_{n,p}(r)\end{pmatrix}\,.
\ee

Similar to bosonic case, we first expand the fermionic fields in terms of Fourier modes. We will not present here the Fourier expansion of the fermionic fields, but we follow closely to the bosonic case e.g.
\bea
&&\Psi^{1}_{t}\=\frac{1}{2}\Psi^{+}_{t;n,p}(r)e^{i(nt+p\th)}+\frac{1}{2}\Psi^{-}_{t;n,p}(r)e^{-i(nt+p\th)}\,,\nn\\
&&\Psi^{1}_{r}\=\frac{i}{2}\Psi^{+}_{r;n,p}(r)e^{i(nt+p\th)}-\frac{i}{2}\Psi^{-}_{r;n,p}(r)e^{-i(nt+p\th)}\,,\nn\\
&&\Psi^{1}_{\theta}\=\frac{1}{2}\Psi^{+}_{\theta;n,p}(r)e^{i(nt+p\th)}+\frac{1}{2}\Psi^{-}_{\theta;n,p}(r)e^{-i(nt+p\th)}\,.
\eea
Then, the fermionic equations of motions are
\be
\mathcal M_{f}E_{f;n,p}^{+}(r)\equiv M_{2f}\p^{2}_{r}E_{f;n,p}^{+}(r)+M_{1f}\p_{r}E_{f;n,p}^{+}(r)+M_{0f}E_{f;n,p}^{+}(r)\=0\,.
\ee
Here $M_{2f,1f,0f}$ are $6\times 6$ matrices which are functions of coordinate $r$, and
\be
E_{f;n,p}^{+}(r)\=\begin{pmatrix}\wt\Psi^{+}_{t;n,p}\\\wt\Psi^{+}_{r;n,p}\\\wt\Psi^{+}_{\theta;n,p}\\c^{+}_{n,p}\\\wt c^{+}_{n,p}\\\Psi^{+}_{n,p}\end{pmatrix}\,.
\ee
Here $\Psi_{\m}\=\wt\Psi_{\m}-D_{\m}c$\,\footnote{Note that the change of the field variable does not involve $\a$. Therefore, one naively expects that the resultant Jacobian will not give any extra $\a$-dependent contribution. We have checked that this naive expectation is indeed correct.}. The explicit form of these matrices are given in Appendix~\ref{appenB}.

\subsubsection*{Greens function}

The Green's function for the bosonic operator is a $4\times 4$ matrix and satisfies the equation
\be
M_{2}\p^{2}_{r}G_{b;n,p}^{+}(r,r')+M_{1}\p_{r}G_{b;n,p}^{+}(r,r')+M_{0}G_{b;n,p}^{+}(r,r')\=\delta(r-r')\,.
\ee
The explicit form of the matrices~$M_{2,1,0}$ are given in the Appendix~\ref{appenB}. One of the simplifications which occur for the choice of the gauge fixing Lagrangian~\eqref{gfLagrangian} is that the equations of motion for the scalar decouples from the equations of motion of the vector field $a_{\m}$. Thus, the bosonic Green's function is block diagonal and has the form~\eqref{bosonicGreenfn.1} for $r<r'$ and~\eqref{bosonicGreenfn.2} for $r>r'$, where in the present case, $G_{1}(r,r')$ (and $G'_{1}(r,r')$) is $3\times 3$ and $G_{2}(r,r')$ (and $G'_{2}(r,r')$) is $1\times 1$ matrix, respectively.

The continuity and discontinuity of the first derivative of the Green's function
\be
G_{b;n,p}^{+}(r,r')\Big|_{r<r'}-G_{b;n,p}^{+}(r,r')\Big|_{r>r'}\=0\,,
\ee
and 
\be
\p_{r}G_{b;n,p}^{+}(r,r')\Big|_{r<r'}-\p_{r}G_{b;n,p}^{+}(r,r')\Big|_{r>r'}\=M^{-1}_{2}\,.
\ee
Similarly, the Green's function for the fermionic operator is a $6\times 6$ matrix which satisfies the similar continuity and discontinuity relations as above.
\subsection{Boundary terms}
Next, we consider the variation of the one loop determinant with respect to the background parameter~$\a$. The variation is
\begin{equation}\label{greendet}
\frac{\delta}{\delta \alpha} \ln Z_{\rm 1-loop} (\a)
= {\rm Tr} [ G_F \frac{\delta}{\delta \alpha} {\cal D}_F(\alpha) ]
- \frac{1}{2} {\rm Tr} [ G_B \frac{\delta}{\delta \alpha} {\cal D}_B(\alpha) ]\,,
\end{equation} 
where~${\cal D}_F(\alpha)$ and~${\cal D}_B(\alpha)$ are fermionic and bosonic kinetic operator, respectively. Following the discussion presented in the Section~\ref{Sec.Var.OneLoop} we find that in the supersymmetric case the variation is a total derivative and is given as
\bea\label{Bdy.Term}
\frac{\delta}{\delta \alpha} \ln Z_{\rm 1-loop} (\a)=-\delta_{\a}(\ln Q^{2})\,\text{tr}B_{1}D^{-1}_{0}CG_{1}(r,r)\Big|^{r=\infty}_{r=0}\,,
\eea
where $G_{1}(r,r')$ is the bosonic Green's function which is constructed out of the solutions of the Equations of motions for the vector field and
\be
B=B_{1}\frac{\p}{\p r}+B_{0},\quad C=C_{1}\frac{\p}{\p r}+C_{0}\,.
\ee
The explicit forms of these matrices are
\be
B_{1}=\begin{pmatrix}0&-\frac{iL}{2}\tanh^{2}r\\-\frac{i}{2}\sinh r&0\\0&\frac{i}{2}\sech^{2}r\end{pmatrix},\quad B_{0}=\begin{pmatrix}\frac{i}{2}L^{2}n\sinh r&-iL\sech^{2}r\tanh r\\-i\sinh r\tanh r&\frac{i}{2\,\cosh^{2}r}(-p+Ln\sinh^{2}r)\\\frac{i}{2}\frac{p}{\sinh r}&-i\sech^{2}r\tanh r\end{pmatrix}\,,
\ee
and
\be
C_{1}=\begin{pmatrix}0&\frac{i}{2}\sinh r&0\\\frac{i}{2}L\tanh^{2}r&0&-\frac{i}{2}\sech^{2}r\end{pmatrix},\,\, C_{0}=\begin{pmatrix}\frac{i}{2}L^{2}n\sinh r&\frac{i}{4\,\cosh r}(3-\cosh 2r)&\frac{ip}{2\sinh r}\\0&\frac{i}{2}(Ln\sinh^{2}r-p)\sech^{2}r&0\end{pmatrix}\,.
\ee
It is not very hard to see that the differential operator~$B$ and~$C$ are adjoint to each other, i.e. $B=C^{\dagger}$. The operator~$D$ is algebraic (not a differential operator) and is given by 
\be
D=\begin{pmatrix}iL\xi(p+L(n-\a))\sinh r&0\\0&-\frac{i}{2}L(p+L(n-\a))\sech r\tanh r\end{pmatrix}\,.
\ee
Note that the matrix operator $B$ (and $C$) are independent of $\a$. Furthermore, the $\a$ dependence in the matrix $D$ is of the form $Q^{2}$ and therefore, the matrix $D$ can be written as $Q^{2}D_{0}$, where $D_{0}$ is independent of $\a$. Thus, it justifies the form of the variation~\eqref{Bdy.Term} where $\delta_{\a}$ acts only on $D$.

It is important to emphasize here that the variation being a total derivative~\eqref{Bdy.Term} depends on the boundary conditions. In fact, the derivation assumes that the fluctuations of fermionic and bosonic fields obey boundary conditions which are consistent with susy. In other words, the fermionic kinetic operator is related to bosonic kinetic operator by a similarity transformations\footnote{One can show that fermionic kinetic operator is $M_{f}=(E^{\dagger})^{-1}\begin{pmatrix}\g_{1}A^{b}_{1}&0&0\\0&\g_{2}A^{b}_{2}&0\\0&0&D\end{pmatrix}E^{-1}$, where $E$ is a $((2k+2)\times(2k+2))$ matrix first order differential operator and $\g_{1}=\frac{1}{Q^{2}}=\frac{1}{\g_{2}}$.} and therefore, the fermionic and bosonic Green's functions are related by similarity transformations. 
We will show below that this is true if $L^{2}>\frac{3}{4}$. When $L^{2}<\frac{3}{4}$, the variation of the one loop determinant will not just be a boundary terms but will also contain bulk terms~\cite{David:2018pex}. 

\subsection{Evaluating boundary terms}

Next, we evaluate the boundary term~\eqref{Bdy.Term}. To evaluate this we just need to determine the action of the first order differential operator~$B_{1}D^{-1}_{0}C$ on the Green's function~$G_{1}(r,r')$ and their asymptotic behaviour. Interestingly, we do not need to know the complete details of the Green's function except it's asymptotic behaviour. As we will see below, this greatly simplifies the computations. The Green's function is constructed from the solutions of the Equations of motions and we will only need to know the asymptotic behaviour of the solutions. 

Now, the Green's function~$G_{1}(r,r')$ satisfies
\be
\mathcal M_{b}\Big|_{X_{0}}G_{1}(r,r')\equiv(m_{b2}\,\p^{2}_{r}+m_{b1}\,\p_{r}+m_{b0})G_{1}(r,r')=\delta(r,r')\,.
\ee
Here $m_{b2,1,0}$ are $3\times 3$ matrices acting on~$X_{0}$ only (which are component of the vector fields). The differential operator $\mathcal M_{b}\Big|_{X_{0}}$ is obtained by projecting the operator $\mathcal M_{b}$ to the vector space $X_{0}$
\be
\mathcal M_{b}\Big|_{X_{0}}=\mathcal P\mathcal M_{b}\mathcal P^{T}\,,
\ee
where
\be
\mathcal P=\begin{pmatrix}1&0&0&0\\0&1&0&0\\0&0&1&0\end{pmatrix}\,.
\ee
\\
In the discussion presented in the Section~\ref{calcbterm}, it turned out to be useful to split the vector space into a rank 1 and rank 2 subspaces. The rank 2 subspace was defined to be the one whose elements are orthogonal to the vector $K$ and the rank 1 whose elements orthogonal to $C_{1}$.
Following the same spirit, we split the vector space $X_{0}$ which we denote by~$V$ into~$V_{1}$ and~$V_{2}$. In the present case, the dimension of the vector space $V$, $V_{1}$ and $V_{2}$ are $3,2$ and $1$, respectively. To define the vector space~$V_{1}$ we need the vector $K$ which is given as (see the Appendix for more details)
\be
K=\frac{1}{p+L(n-\a)}\begin{pmatrix}L&0&1\end{pmatrix}.
\ee
A typical vector in $V_{1}$ has the form
\be
v_{1}=\begin{pmatrix}-\frac{x_{1}}{L}\\x_{2}\\x_{1}\end{pmatrix},\quad x_{1,2}\in\mathbb R
\ee
and that belonging to the vector space $V_{2}$ has the form
\be
v_{2}=x\begin{pmatrix}1\\0\\L\sinh^{2}r\end{pmatrix},\quad x\in\mathbb R
\ee
In order to simplify the computations, we change the basis of the vector space~$V$ such that the first two non zero component belongs to the vector space~$V_{1}$ and the 3rd non zero component belongs to the vector space~$V_{2}$.
That is given a vector $v\in V$, we define a vector $\wt v$ as $v=J\,\wt v$ such that for $\wt v=\begin{pmatrix}c_{1}\\c_{2}\\0\end{pmatrix}$, for $c_{1,2}\in\mathbb R$, the corresponding $v\in V_{1}$ and for $\wt v=\begin{pmatrix}0\\0\\c_{3}\end{pmatrix}$, for $c_{3}\in\mathbb R$, the corresponding $v\in V_{2}$. 
It turns out that there is no unique choice of $J$ (different $J$'s are related to each other by rotation in $V_{1}$ space) and one convenient choice is
\be
J=\begin{pmatrix}0&\frac{2}{L}\tanh r&\frac{2}{L}\sech r\\2&0&0\\0&-2\tanh r&2\sinh r\,\tanh r \end{pmatrix}\,.
\ee
Subsequently, the corresponding matrix operator acting on the elements of the vector space $\wt V$ is related to the original operator by similarity transformations as
\be
m_{b2p,b1p,b0p}=J^{T}m_{b2,b1,b0}J\,. 
\ee
\subsubsection*{Asymptotic behaviour of differential operator:}

As we found earlier in~\eqref{Bdy.Term} that to evaluate the variation of the one loop determinant, we just need to know the asymptotic behaviour of the Green's function. Now, the Green's functions are constructed out of the solutions of the Equations of motion. Thus for our purposes to evaluate the boundary terms~\eqref{Bdy.Term}, the global form of the solutions are not necessary rather its asymptotic form will suffice. Furthermore, we argued there that the contributions to the boundary terms only come from the space of the solutions belonging to the vector space~$V_{1}$. Thus we need to construct the Green's function restricted to the vector space~$V_{1}$ i.e.
\be
G_{1}(r,r')\Big|_{V_{1}}=PG_{1}(r,r')P^{T}\,,
\ee
where the projection operator is
\be
P=\begin{pmatrix}1&0&0\\0&1&0\end{pmatrix}\,.
\ee

To obtain the asymptotic form of the solutions, we need to analyse the asymptotic behaviour of the kinetic operator near $r=0$ and $r=\infty$. Near $r\rightarrow 0$, the leading contributions to matrix coefficients of the 2nd order differential operator are
\bea
&&\lim_{r\rightarrow 0}m_{b2p}=\frac{r}{L^{2}}\begin{pmatrix}\frac{1}{\xi}&0&0\\0&-2&0\\0&0&-2\end{pmatrix}+\mathcal O(r^{3}),\,\, \lim_{r\rightarrow 0}m_{b1p}=\frac{1}{\xi L^{2}}\begin{pmatrix}1&-p(1+2\xi)&0\\p(1+2\xi)&-2\xi&0\\0&0&-2\xi\end{pmatrix}+\mathcal O(r)\,,\nn\\
&&\lim_{r\rightarrow 0}m_{b0p}=\frac{1}{L^{2}r\xi}\begin{pmatrix}2p^{2}\xi-1&p(1-2\xi)&0\\p(1-2\xi)&2\xi-p^{2}&0\\0&0&2p^{2}\xi\end{pmatrix}+\mathcal O(1)\,.
\eea
On the other hand near $r\rightarrow\infty$, the leading behaviour of the differential operator is  
\bea
&&\lim_{r\rightarrow \infty}m_{b2p}=\frac{1}{uL^{2}}\begin{pmatrix}\frac{1}{2\xi}&0&0\\0&-1&0\\0&0&-1\end{pmatrix}+\mathcal O(1),\,\, \lim_{r\rightarrow \infty}m_{b1p}=\frac{1}{uL^{2}}\begin{pmatrix}\frac{1}{2\xi}&\frac{Ln(1+2\xi)}{2\xi}&0\\-\frac{Ln(1+2\xi)}{2\xi}&-1&0\\0&0&-1\end{pmatrix}+\mathcal O(1)\,,\nn\\
&&\lim_{r\rightarrow \infty}m_{b0p}=\frac{1}{uL}\begin{pmatrix}Ln^{2}-\frac{1}{L\xi}&\frac{n}{\xi}&0\\\frac{n(1-2\xi)}{2\xi}&-\frac{Ln^{2}}{2\xi}&0\\0&0&Ln^{2}\end{pmatrix}+\mathcal O(1)\,.
\eea
Here $u=e^{-r}$.

It is important to observe that the second order differential operator $J^{T}BD^{-1}CJ\Big|_{V_{1}}$ has the same asymptotic behaviour as the differential operator 

\ndt{\bf Solutions near $r\rightarrow 0$:}
The asymptotic behaviour of the solution near $r\rightarrow 0$ is controlled by the integer $p$ and is independent of $n$. Solving the Equations of motion near $r\rightarrow 0$ we find that, for $p>0$, there are 3 smooth solutions which are 
\bea\label{Sols.r=0.1}
&&s_{1p}(r)=r^{p-1}\begin{pmatrix}1\\1\\0\end{pmatrix}\in V_{1},\qquad
s_{2p}(r)=r^{p+1}\begin{pmatrix}\frac{p(1+2\xi)+4\xi}{2+p(1+2\xi)}\\1\\0\end{pmatrix}\in V_{1}\,,\nn\\
&&s_{3p}(r)=r^{p}\begin{pmatrix}0\\0\\1\end{pmatrix}\in V_{2}\,,
\eea
and 3 singular solutions which are 
\bea\label{Sols.r=0.2}
&&s_{4p}(r)=r^{-p-1}\begin{pmatrix}1\\-1\\0\end{pmatrix}\in V_{1},\qquad s_{5p}(r)=r^{-p+1}\begin{pmatrix}1\\-\frac{p(1+2\xi)-2}{p(1+2\xi)-4\xi}\\0\end{pmatrix}\in V_{1}\,,\nn\\
&&s_{6p}(r)=r^{-p}\begin{pmatrix}0\\0\\1\end{pmatrix}\in V_{2}\,.
\eea
For $p<0$, the solutions $s_{4p,5p,6p}$ are smooth and $s_{1p,2p,3p}$ are singular. 

For the case of $p=0$, we see that $s_{3}(r)$ and $s_{6}(r)$ are degenerate. Solving next to leading order we find two linearly independent solutions and are given by
\be
s_{30}(r)=\begin{pmatrix}0\\0\\1\end{pmatrix}\in V_{2},\quad s_{60}(r)=\ln r\begin{pmatrix}0\\0\\1\end{pmatrix}\in V_{2}\,.
\ee
Thus, for $p=0$, the solutions which are smooth are $s_{20,30,50}$ whereas $s_{10,40,60}$ are singular near $r\rightarrow 0$, where $s_{20,50}$ and $s_{10,40}$ are obtained by putting $p=0$ in $s_{2p,5p}$ and $s_{1p,4p}$, respectively.

Since $\wt\Psi_{\m}$ satisfies the same Equations of motion as the vector field, therefore, the smooth solutions for the vector field are also smooth for $\wt\Psi_{\m}$. Near $r\rightarrow 0$ behaviour of the solution for $(\wt c,\Psi)$ is obtained from $\wt\Psi_{\m}$ as
\be\label{PsiToPsimu}
\begin{pmatrix}\wt c^{+}_{n,p}\\\Psi^{+}_{n,p}\end{pmatrix}=-D^{-1}C\begin{pmatrix}\wt\Psi^{+}_{t;n,p}\\\wt\Psi^{+}_{r;n,p}\\\wt\Psi^{+}_{\theta;n,p}\end{pmatrix}\,.
\ee
Using the solutions given in~\eqref{Sols.r=0.1} and~\eqref{Sols.r=0.2} for $p> 0$, we find that $s_{1p}(r),s_{2p}(r)$ and $s_{3p}(r)$ also give rise smooth solutions for $\wt c$ and $\Psi$. For example when $s_{1p}(r),s_{2p}(r)$ and $s_{3p}(r)$ are acted upon by the differential operator $-D^{-1}C$, we get near $r\rightarrow 0$
\be
-D^{-1}C s_{1p}(r)\sim\begin{pmatrix}c_{1}\\c_{2}\end{pmatrix}r^{p},\quad -D^{-1}C s_{2p}(r)\sim c_{3}\begin{pmatrix}-1\\1\end{pmatrix}r^{p},\quad -D^{-1}C s_{3p}(r)\sim\begin{pmatrix}c_{4}\\c_{5}\end{pmatrix}r^{p}\,.
\ee
Here $c_{i}$'s are constants. Thus, for $p> 0$ above are smooth solutions for $\wt c$ and $\Psi$. Similarly, it is not difficult to see that $s_{4p,5p,6p}$ do not give smooth solution near $r\rightarrow 0$. For $p<0$, the smooth solutions for $\wt c$ and $\Psi$ are obtained from $s_{4p,5p,6p}$ whereas $s_{1p,2p,3p}$ give rise singular solutions. For $p=0$, $s_{i0}$ for $i=1,..5$ give smooth solutions for $\wt c$ and $\Psi$. Thus, for the fermionic system $(\wt\Psi_{\m},\wt c,\Psi)$, for $p=0$, the smooth solutions are $s_{20,30,50}$ whereas $s_{10,40,60}$ are singular.

\ndt{\bf Solutions near $r\rightarrow \infty$:}
Next, we determine the asymptotic behaviour of the solutions near $r\rightarrow \infty$. The asymptotic behaviour of the solution near $r\rightarrow \infty$ is controlled by the integer $n$ and is independent of $p$. 
We find that for $L^{2}n^{2}>\frac{3}{4}$, following are the asymptotic behaviour of normalizable solutions (normalizability conditions for the component of gauge field are given in~\eqref{NormConditions.1})
\bea\label{AsymptoticSols.1}
&&\wt s_{1n}=\begin{pmatrix}c(n)\\1\\0\end{pmatrix}e^{-\frac{r}{2}(3+\sqrt{1+4L^{2}n^{2}})}\in V_{1},\quad \wt s_{2n}=\begin{pmatrix}\wt c(n)\\1\\0\end{pmatrix}e^{-\frac{r}{2}(-1+\sqrt{1+4L^{2}n^{2}})}\in V_{1}\,,\nn\\
&&\wt s_{3n}=\begin{pmatrix}0\\0\\1\end{pmatrix}e^{-\frac{r}{2}(1+\sqrt{1+4L^{2}n^{2}})}\in V_{2}\,.
\eea
Here $c(n)=\frac{Ln\Big(-1+6\xi+\sqrt{1+4L^{2}n^{2}}\,(1+2\xi)\Big)}{2\Big(-1+\sqrt{1+4L^{2}n^{2}}+L^{2}n^{2}(1+2\xi)\Big)}$ and $\wt c(n)=-\frac{Ln\Big(-5-2\xi+\sqrt{1+4L^{2}n^{2}}\,(1+2\xi)\Big)}{2\Big(1+\sqrt{1+4L^{2}n^{2}}-L^{2}n^{2}(1+2\xi)\Big)}$. 
\vspace{0.2cm}

The asymptotic behaviour of solutions which are not normalizable are
\bea\label{AsymptoticSols.2}
&&\wt s_{4n}=\begin{pmatrix}c_{1}(n)\\1\\0\end{pmatrix}e^{-\frac{r}{2}(-1-\sqrt{1+4L^{2}n^{2}})}\in V_{1},\quad \wt s_{5n}=\begin{pmatrix}\wt c_{1}(n)\\1\\0\end{pmatrix}e^{-\frac{r}{2}(3-\sqrt{1+4L^{2}n^{2}})}\in V_{1}\,,\nn\\
&&\wt s_{6n}=\begin{pmatrix}0\\0\\1\end{pmatrix}e^{-\frac{r}{2}(1-\sqrt{1+4L^{2}n^{2}})}\in V_{2}\,.
\eea
Here $c_{1}(n)=-\frac{Ln\Big(5+2\xi+\sqrt{1+4L^{2}n^{2}}\,(1+2\xi)\Big)}{2\Big(-1+\sqrt{1+4L^{2}n^{2}}+L^{2}n^{2}(1+2\xi)\Big)}$ and $\wt c_{1}(n)=\frac{Ln\Big(1-6\xi+\sqrt{1+4L^{2}n^{2}}\,(1+2\xi)\Big)}{2\Big(1+\sqrt{1+4L^{2}n^{2}}-L^{2}n^{2}(1+2\xi)\Big)}$.
\vspace{0.2 cm}

However, for $0<L^{2}n^{2}<\frac{3}{4}$, we find that the normalizable solutions are $\wt s_{1n}(r),\wt s_{3n}(r)$ and $\wt s_{5n}(r)$ and non normalizable solutions are $\wt s_{2n}(r),\wt s_{4n}(r)$ and $\wt s_{6n}(r)$. 
The solution with~$n=0$ will play an important role for later analysis, we present here their explicit form.
For $n=0$, the asymptotic behaviour of normalizable solutions are
\bea\label{n=0;normalizable}
&&\wt s_{10}(r)=\begin{pmatrix}1\\0\\0\end{pmatrix}e^{-2r}\in V_{1},\quad \wt s_{30}(r)=\begin{pmatrix}0\\0\\1\end{pmatrix}e^{-r}\in V_{2}\,,\nn\\
&&\wt s_{50}(r)=\begin{pmatrix}0\\1\\0\end{pmatrix}e^{-r}\in V_{1}.
\eea
and the asymptotic behaviour of the non normalizable solutions are
\bea\label{n=0;nonnormalizable}
&&\wt s_{20}(r)=\begin{pmatrix}0\\1\\0\end{pmatrix}\in V_{1},\quad \wt s_{40}(r)=\begin{pmatrix}1\\0\\0\end{pmatrix}e^{r}\in V_{1}\,,\nn\\
&&\wt s_{60}(r)=\begin{pmatrix}0\\0\\1\end{pmatrix}\in V_{2}\,.
\eea
Now, we discuss asymptotic behaviour of solutions belonging to fermionic system $(\wt\Psi_{\m},c,\wt c,\Psi)$. Since $\wt\Psi_{\m}$ satisfies the same Equation of motion as the vector field, the solutions of vector field are also solutions for the fermion $\wt\Psi_{\m}$. However, fields $(\wt\Psi_{\m},c,\wt c,\Psi)$ have different normalizabilty conditions, see~\eqref{NormConditions.2} and ~\eqref{NormConditions.3}, and therefore, we need to reanalyse which of the solutions among the set of solutions obtained above are normalizable and non normalizable, respectively for fermions. Before going to analyse the above solutions for fermions, it is important to mention a few comments about the Equation of motion satisfied by~$c$. From susy algebra~\eqref{QhatTransf}, we see that if we replace $c$ by $\cosh r\,\hat c-\frac{1}{Q^{2}}K^{\m}\wt\Psi_{\m}$, then $\hat c$ satisfies the same equation as $\sigma$. Solving the Equation of motion for $\sigma$ we find that there are 2 solutions with asymptotic behaviour near $r=\infty$
\be
\sigma_{n,p}\sim A_{1} e^{-\frac{r}{2}(1-\sqrt{1+4L^{2}n^{2}})}+A_{2}e^{-\frac{r}{2}(1+\sqrt{1+4L^{2}n^{2}})}\,.
\ee
The normalizablity condition on $\sigma$ requires us to choose the second solution. Since~$\hat c_{n,p}$ satisfies the same equation as~$\sigma_{n,p}$, we have the same asymptotic behaviour for~$\hat c_{n,p}$. Thus, it is easy to see that for the ghost~$c_{n,p}$, it is only the 2nd solution (labelled by~$A_{2}$) which will give admissible asymptotic behaviour. Furthermore, given the asymptotic behaviour of the solutions~$\wt\Psi_{\m}$, the asymptotic behaviour of $(\wt c,\Psi )$ is obtained by using~\eqref{PsiToPsimu}. Now, we will tabulate these solutions indicating whether they are normalizable (marked by $\checkmark$) or nonnormalizable (marked by $\text{\sffamily X}$).
\begin{table}
\begin{center}
\begin{tabular}{ |c|c|c|c|c| } 
 \hline
 Solutions & $\wt\Psi_{\m}$ & $c=\cosh r\,\hat c-\frac{1}{Q^{2}}K^{\m}\wt\Psi_{\m}$&$\wt c$&$\Psi$ \\ 
 \hline 
 $\wt s_{1n}(r)$  &  $\checkmark\,\,\forall\,n\neq 0$ & $\checkmark\,\,\forall\,n\neq 0$&$\checkmark\,\,\forall\,n\neq 0$& $\checkmark\,\,\forall\,n\neq 0$ \\ 
 \hline
 $\wt s_{2n}(r)$  &  $\checkmark\,\,\forall\,n\neq 0$ & $\checkmark\,\,\forall\,n\neq 0$&$\checkmark\,\,\forall\,n\neq 0$& $\checkmark\,\,\forall\,n\neq 0$  \\ 
 \hline
$\wt s_{3n}(r)$  &  $\checkmark\,\,\forall\,n\neq 0$ & $\checkmark\,\,\forall\,n\neq 0$&$\checkmark\,\,\forall\,n\neq 0$& $\checkmark\,\,\forall\,n\neq 0$  \\ 
\hline
$\wt s_{4n}(r)$  &  $\text{\sffamily X}\,\,\forall\,n\neq 0$ & $\text{\sffamily X}\,\,\forall\,n\neq 0$&$\text{\sffamily X}\,\,\forall\,n\neq 0$& $\text{\sffamily X}\,\,\forall\,n\neq 0$  \\ 
\hline
$\wt s_{5n}(r)$  &  $\checkmark\,\,0<\forall\,L^{2}n^{2}<2$ & $\checkmark\,\ 0<\forall\,L^{2}n^{2}\leq 2$&$\text{\sffamily X}\,\,\forall\,n\neq 0$& $\text{\sffamily X}\,\,\forall\,n\neq 0$  \\ 
\hline
$\wt s_{6n}(r)$  &  $\text{\sffamily X}\,\,\forall\,n\neq 0$ & $\text{\sffamily X}\,\,\forall\,n\neq 0$&$\text{\sffamily X}\,\,\forall\,n\neq 0$& $\text{\sffamily X}\,\,\forall\,n\neq 0$  \\ 
\hline
\end{tabular}
\end{center}
\caption{Summary of acceptable and non acceptable solutions for fermionic fields.}
\end{table}
Looking at the table, we see that the solutions for bosonic and fermionic fields are consistent with supersymmetry only for $L^{2}n^{2}>\frac{3}{4}$. For the range $0<L^{2}n^{2}<\frac{3}{4}$, we find that~$\wt s_{1n}(r),\wt s_{3n}(r)$ and $\wt s_{5n}(r)$ are normalizable for the gauge field whereas~$\wt s_{1n}(r),\wt s_{2n}(r)$ and $\wt s_{3n}(r)$ are normalizable for fermionic fields. Thus, there is a mismatch of the space of allowed solutions for fermionic and bosonic fields. In this situation, the Green's function for the bosonic field is not related to that of the fermionic field and, therefore, for the modes lying in the interval, $0<L^{2}n^{2}<\frac{3}{4}$, the variation of the one loop determinant will not be just a boundary terms but will also include bulk terms. To determine the explicit expression for the bulk term we need to know the global form of the solutions and not just the asymptotic behaviour. This is a much more harder problem in the present case where we do not have the global form of the solution. To avoid this, we assume that $L^{2}>\frac{3}{4}$. With this there are no modes lying in the interval $0<L^{2}n^{2}<\frac{3}{4}$.

Now, we will discuss the case of $n=0$. In this case the analysis is slightly subtle and needs a separate discussion.

\ndt{\bf Case: $n=0$}\\
The acceptable solutions for the bosonic fields are given in~\eqref{n=0;normalizable}. Next we need to analyze whether these solutions give rise acceptable solutions to fermionic fields. In this case it turns out that $\wt s_{10}$ and $\wt s_{30}$ give rise normalizable solutions, while $\wt s_{40}$ and $\wt s_{60}$ give rise nonnormalizable solutions to the fermionic fields. The asymptotic behaviour of the solutions $\wt s_{50}$ and $\wt s_{20}$ are subtle for fermionic fields. For these solutions, we find the following: For the solutions $\wt s_{50}$, the asymptotic behaviour of fermionic fields as $r\rightarrow\infty$ are
\be
\wt s_{50} :\quad \wt\Psi_{t}\sim e^{-r},\,\, \wt\Psi_{r}\sim 0,\,\, \wt\Psi_{\theta}\sim e^{-r},\quad \wt c\sim e^{-3r},\,\,\Psi\sim \mathcal O(1),\,\,c\sim \mathcal O(1)\,,
\ee
whereas for the solutions $\wt s_{20}$, the asymptotic behaviour of fermionic fields as $r\rightarrow\infty$ are
\be
\wt s_{20} :\quad \wt\Psi_{t}\sim \mathcal O(1),\,\, \wt\Psi_{r}\sim 0,\,\, \wt\Psi_{\theta}\sim \mathcal O(1),\quad \wt c\sim e^{-r},\,\,\Psi\sim e^{-r},\,\,c\sim \mathcal O(1)\,.
\ee
Comparing these asymptotic behaviour with the boundary conditions~\eqref{NormConditions.2} and the boundary condition on $\wt c$, one would naively declare both the above solutions to be non normalizable. But this would amount to non existence of Green's function. The requirement of the existence of the Green's function forces us to declare one of these solution to be normalizable and other to be nonnormalizable. Thus, for the case of  $n=0$ and $p\neq0$ we have two choices : \\
\ndt 1) We declare that $\wt s_{50}$ is normalizable and $\wt s_{20}$ is nonnormalizable which would correspond to preserving supersymmetry, or \\
\ndt 2) We declare $\wt s_{20}$ to be normalizable and $\wt s_{50}$ nonnormalizable, then this would break the supersymmetry.

Making either of the choices requires to modify (although minimally) the boundary conditions we started with. Since we are only interested in the boundary terms, which is the case when the allowed modes are also consistent with supersymmetry, we choose the option 1. It is definitely worth to try with option 2, but in this case we also need to calculate the bulk term (because for this choice we do not have supersymmetric cancellation) which is beyond the scope of the present paper.
To allow the option 1, we modify the boundary conditions~\eqref{NormConditions.2} which would amount to following asymptotic behaviour
\be\label{NormConditions.3}
\Psi_{t}\rightarrow 0,\quad \Psi_{r}\rightarrow 0,\quad e^{-r}\Psi_{\theta}\rightarrow 0,\quad \text{and}\quad \Psi\rightarrow \mathcal O(1)\,.
\ee
Note that this choice does not change the analysis presented above for the case $n\neq 0$.
\subsection{Variation of the one loop partition function}

As it was shown in the Section~\ref{calcbterm} that to determine the boundary contribution we just need to know the dimension of the kernel of the operator $C|_{V_{1}}$ i.e $\ell$ and $\ell'$ near $r\rightarrow 0$ and $r\rightarrow \infty$, respectively. Theses dimensions of the kernel of the operator $C|_{V_{1}}$ depends on the value of $(n,p)$. We split the evaluation of the boundary term in following 4 different cases:

\ndt{\bf Case: $p\neq 0,n\neq 0$}

We start with the computation of the boundary term near $r\rightarrow 0$. As we found earlier, the asymptotic behaviour of the solutions in this limit depends only on the value of $p$ and are independent of $n$. The solutions which are admissible near $r\rightarrow 0$ for $p>0$ and $p<0$ are $s_{1p}(r)$ and $s_{2p}(r)$, and $s_{4p}(r)$ and $s_{5p}(r)$, respectively.  However, it is only $s_{1p}(r)$ ( $s_{4p}(r)$) belongs to the kernel of $C|_{V_{1}}$ for $p>0$ $(p<0)$ i.e.
\be
\lim_{r\rightarrow 0}C|_{V_{1}}s_{1p}(r)=0,\quad \text{for}\,\,p>0\,,
\ee
and
\be
\lim_{r\rightarrow 0}C|_{V_{1}}s_{4p}(r)=0,\quad \text{for}\,\,p<0\,.
\ee
Thus the dimension of the kernel, $\ell$, for $p\neq0$ is 1. 

Near $r\rightarrow \infty$, the admissible solutions are $s_{1n}(r)$ and $s_{2n}(r)$. However, the solution which belongs to the kernel of $C|_{V_{1}}$ is $s_{2n}(r)$ i.e.
\be
\lim_{r\rightarrow \infty}C|_{V_{1}}s_{2n}(r)=0\,.
\ee
Thus, we have $\ell'=1$. Therefore, the boundary contribution for the case $n\neq0$ and $p\neq0$ is
\be
\text{BT}=(k-\ell-\ell')=2-1-1=0\,.
\ee

\ndt{\bf Case: $p\neq 0,n= 0$}

Since the asymptotic behaviour near $r\rightarrow 0$ for $p\neq 0$ does not depend on $n$, the dimension of the kernel, $\ell$, remains same as before and is equal to 1. However, for $n=0$ we find that there no normalizable modes in \eqref{n=0;normalizable} which belongs to the kernel of $C|_{V_{1}}$. Thus in this case we have $\ell'=0$. Therefore, the boundary contribution for the case $n=0$ and $p\neq0$ is
\be
\text{BT}=(k-\ell-\ell')=2-1=1\,.
\ee
\ndt{\bf Case: $n\neq 0,p= 0$}

Since the asymptotic behaviour near $r\rightarrow \infty$ for $n\neq 0$ does not depend on $p$, the dimension of the kernel, $\ell'$, remains same as before and is equal to 1. However, for $p=0$ we find that there no smooth modes which belongs to the kernel of $C|_{V_{1}}$. Thus in this case we have $\ell=0$. Therefore, the boundary contribution for the case $p=0$ and $n\neq0$ is
\be
\text{BT}=(k-\ell-\ell')=2-1=1\,.
\ee
\ndt{\bf Case: $n=p= 0$}

Following the discussion of $n=0,p\neq 0$ and $p=0,n\neq 0$ cases we find that dimensions of the kernel of $C|_{V_{1}}$ in the case of $n=p=0$ are $\ell=\ell'=0$. Thus, its contribution to the boundary term is
\be\label{BT00Case}
\text{BT}=2\,.
\ee 
It was observed in~\cite{David:2016onq} that this contribution to the index comes precisely from the zero modes of the ghost fields which were given by globally constants mode for ghost $c$ and anti ghost $\wt c$. Since the determinant are computed over non zero modes, we did not include the contribution of these zero modes. 

We also observe this fact in our present computation. First zero mode corresponds scalar fluctuations parallel to the localization background i.e.
\be
\hat\sigma\=\frac{\vec A}{\cosh r}\,,\qquad \vec A=\text{constant Lie algebra element}\,.
\ee
The supersymmetric partner of the above zero mode is the constant ghost mode $c=\vec A$ (it can be seen following~\eqref{SusyTransf.ofGhost}). 

As we discussed near \eqref{ZeroModeTildeC}, the second zero mode corresponds to
\be
\wt c\= \frac{\vec \m}{\cosh^{2}r},\qquad b=\frac{\vec\m_{Q}}{\cosh^{2}r},\quad \text{where}\,\,\,\,Q\vec\m_{Q}=i[\Lambda^{(0)},\vec\m]\,,
\ee
where~$\vec\mu$ and $\vec\m_{Q}$ are Grassmann odd and even constant Lie algebra element, respectively. The ghost Lagrangian involves mass like terms
\be
\frac{\xi}{\cosh^{2}r}\Big(\text{tr} \vec\m_{Q}^{2} + 2\sum_{\vec\rho>0}\rho.\a\,\m_{-\rho}\m_{\rho} \Big)\subset\mathcal L_{\text{g.f.}}\,.
\ee
Here $\vec\rho$ is a root of the Lie algebra. The first terms comes from $\text{tr}\,b^{2}$ and the second term comes from $\text{tr}\,\wt c\,[\Lambda_{0},\wt c]$. 
Integrating over this mode and then calculating its variation with respect to~$\a$ gives rise $\text{BT}=1$. Since this contribution is a zero mode contribution and we are computing determinant over non zero mode, we subtract $1$ from \eqref{BT00Case}.

To treat the zero mode $c=\vec A$, we need to use the method of generalized Green's function. In this method, the Green's function equation is modified by a zero mode projector. Because of the presence of the zero mode projector, the variation of the one loop determinant, after performing integration by parts, now gives a boundary terms together with an extra bulk term proportional to number of zero modes (coming from the zero mode projector). We will not present the details of this calculation here. However, we find that following this method we get an extra $-1$ in \eqref{BT00Case}. Thus taking into account all zero mode we get $\text{BT}=0$ for the case $n=0=p$.

Thus, collecting all the above results we find that for $L^{2}>\frac{3}{4}$, the variation of the one loop determinant~\eqref{boundaryterm.2} is (for a general compact gauge group)

\be\label{Result1.VarOneLoop}
\delta_{\vec\a}\ln Z\=-\frac{i}{2}\sum_{n\neq0}\frac{\vec \rho}{n-i\rho\cdot\a}-\frac{i}{2}\sum_{p\neq0}\frac{\vec \rho}{\frac{p}{L}-i\rho\cdot\a}\,.
\ee
Integrating with respect to $\a$, we obtain
\be\label{OneLoopInfiniteProd.}
Z_{1-\text{loop}}(\a)\=\prod_{\rho}\sqrt{\prod_{n\neq 0}(n-i\rho\cdot\a)\prod_{p\neq 0}(\frac{p}{L}-i\rho\cdot\a)}\,.
\ee
which is the result obtained in~\cite{David:2016onq}.
Thus, the partition function of a Chern Simons theory with level $k$ and the gauge group $G$ of rank $r$ is
\be
Z\=\int_{\mathbb R^{r}} d\a\,\exp(-\pi iL k\text{Tr}\a^{2})\prod_{\rho>0}\sinh(\pi \rho\cdot\a)\sinh(\pi L\rho\cdot\a)\,.
\ee
In the above, the integration variable $\a$ is valued in the Cartan of the Lie algebra of the gauge group $G$.
Furthermore, we have also included the contribution of the Vandermonde determinant (it is the Jacobian coming from rotating any constant Lie algebra element to an element in the Cartan) to convert the infinite product~\eqref{OneLoopInfiniteProd.} to the product of hyperbolic function.

If we also include matter field which consist of $N_{f}$ chiral multiplets transforming in some representation $\mathcal R_{i}$, where $i=1,...,N_{f}$, of the gauge group, then the partition function of a Chern Simons theory matter theory is given by
\be\label{PartitionFunct.andMatter}
Z=\int_{\mathbb R^{r}} d\a\,\exp(-\pi iL k\text{Tr}\a^{2})\prod_{\rho>0}\sinh(\pi \rho\cdot\a)\sinh(\pi L\rho\cdot\a)\,\prod_{i=1}^{N_{f}}\prod_{\rho}Z^{1-\text{loop}}_{\text{matter}}(\mathcal R_{i};\a)\,,
\ee
where~$Z^{1-\text{loop}}_{\text{matter}}(\mathcal R;\a)$ is the one loop determinant of the chiral multiplet in the representation~$\mathcal R$. It was demonstrated  in~\cite{David:2018pex} using the  Greens function 
method, 
to compute the one loop determinant~$Z^{1-\text{loop}}_{\text{matter}}(\mathcal R;\a)$, that arises
in Localization 
 depends on the choice of $Q$-exact action. In particular, this difference arose for the modes in the interval~$\frac{\Delta-1}{2L}<n<\frac{\Delta}{2L}$. 
\section{Level-rank duality on $AdS_2\times S^1$} \label{s-duality}
In this section, we will discuss one of the implications of the result obtained in the last section for Chern Simons matter theory. We find that for the cases when there are no bulk terms in the partition function i.e. when the normalizable boundary conditions are consistent with supersymmetry (which is the case when $L^{2}>\frac{3}{4}$, and there are no integers in the interval~$\frac{\Delta-1}{2L}<n<\frac{\Delta}{2L}$), the partition function respects 3 dimension level-rank duality. We will consider here the example of $U(N)$ Chern Simons theory coupled to $N_{f}$ hypermultiplets in fundamental (i.e. $N_{f}$ chiral in fundamental and $N_{f}$ chiral multiplet in anti fundamental) with R-charge $\Delta$.
In this case the statement of level-rank duality is
\be\label{LevelRankDuality}
N_{f}\,\,\text{hypermultiplet coupled to}\,\, U(N_{c})_{k} \iff N_{f}\,\,\text{hypermultiplet coupled to}\,\, U(|k|+N_{f}-N_{c})_{-k}
\ee
We will find that this duality also holds true for $U(N)$ Chern Simons theory coupled to $N_{f}$ hypermultiplets in fundamental on $AdS_{2}\times S^{1}$. 

Without loss of generality we will assume that the ratio of the size $L=1$. However, one can generalize the discussion below for any value of $L$ such that $L^{2}>\frac{3}{4}$. Also for the presentation, we will also consider three different cases for which there are no integer in the interval $\frac{\Delta-1}{2L}<n<\frac{\Delta}{2L}$ : 1) with no matter fields ($N_{f}=0$), 2) $N_{f}$ hypermultiplets in fundamental with R-charge $\Delta=0$, and 3) $N_{f}$ hypermultiplets in fundamental with R-charge $\Delta=1$.

\ndt{\bf Case: $N_{f}=0$} 

In this case the partition function \eqref{PartitionFunct.andMatter} reduces to the partition function of a pure Chern Simons theory which is
\be
Z=\int_{\mathbb R^{r}} d\a\,\exp(-\pi i k\text{Tr}\a^{2})\prod_{\rho>0}\sinh^{2}(\pi \rho\cdot\a)\,.
\ee
This partition function is exactly same as the partition function of $U(N)_{k}$ Chern Simons theory on S$^{3}$. 

\ndt{\bf Case: $N_{f}$ Hypermultiplets with $\Delta=0$} 

For a chiral multiplet with R-charge $\Delta=0$, the one loop contribution to the partition function \eqref{PartitionFunct.andMatter} is given  by
\be
\ln Z^{1-\text{loop}}_{\text{matter}}=\sum_{p>0,n\geq 0}\ln(p+n+i\rho(\a))-\sum_{p\leq 0,n<0}\ln(-p-n-i\rho(\a))\,.
\ee
Thus for a given hypermultiplet with R-charge $\Delta=0$, the one loop contribution to the partition function is
\bea
\ln Z^{1-\text{loop}}_{\text{hyper-matter}}&=&\sum_{p>0,n\geq 0}\ln(p+n+i\rho(\a))-\sum_{p\leq 0,n<0}\ln(-p-n-i\rho(\a))\nn\\
&&+\sum_{p>0,n\geq 0}\ln(p+n-i\rho(\a))-\sum_{p\leq 0,n<0}\ln(-p-n+i\rho(\a))=0\,.
\eea
Therefore, in this case there are no contribution to the partition function from the fields in the matter sector. Thus the partition function of $U(N)_{k}$ Chern Simons theory coupled to $N_{f}$ hypermultiplet with R-charge $\Delta=0$ is equal to the partition function of $U(N)_{k}$ Chern Simons theory. Note that this is same as the partition function of $U(N)_{k}$ Chern Simons theory coupled to $N_{f}$ hypermultiplet on S$^{3}$ but with R-charge $\Delta=1$.

\ndt{\bf Case: $N_{f}$ Hypermultiplets with $\Delta=1$}

For a chiral multiplet with R-charge $\Delta=1$, the one loop contribution to the partition function \eqref{PartitionFunct.andMatter} is given  by
\be
\ln Z^{1-\text{loop}}_{\text{matter}}=\sum_{p>0,n\geq 1}\ln(p+n+i\rho(\a)-\frac{1}{2})-\sum_{p\leq 0,n\leq0}\ln(-p-n-i\rho(\a)+\frac{1}{2})\,.
\ee
Thus for a given hypermultiplet with R-charge $\Delta=1$, the one loop contribution to the partition function is
\bea
Z^{1-\text{loop}}_{\text{hyper-matter}}=\frac{\prod_{r= 1}(r+i\rho(\a)+\frac{1}{2})^{r}(r-i\rho(\a)+\frac{1}{2})^{r}}{\prod_{r=1}(r-i\rho(\a)-\frac{1}{2})^{r}(r+i\rho(\a)-\frac{1}{2})^{r}}=\frac{1}{2\cosh\pi\rho(\a)}\,.
\eea
Therefore, the partition function of $U(N)_{k}$ Chern Simons matter theory coupled to $N_{f}$ fundamental hypermultiplet with R-charge $\Delta=1$ is
\be
Z=\int_{\mathbb R^{r}} d\a\,\exp(-\pi iL k\text{Tr}\a^{2})\prod_{\rho>0}\sinh(\pi \rho\cdot\a)\sinh(\pi L\rho\cdot\a)\prod_{\rho}\Big(\frac{1}{2\cosh\pi\rho(\a)}\Big)^{N_{f}}\,.
\ee 
The above is the partition function of $U(N)_{k}$ Chern Simons matter theory coupled to $N_{f}$ fundamental hypermultiplet on S$^{3}$ with R-charge $\Delta=\frac{1}{2}$. It is known that this partition function respects the duality~\eqref{LevelRankDuality}.
\section{Conclusions} \label{discuss}

In this paper we  have developed the method of Greens function introduced in \cite{David:2018pex} 
to evaluate 
one loop determinants that occur in localization of supersymmetric  field theories on $AdS$ spaces. 
The method requires the theory to have at least ${\cal N}=2$ supersymmetry in the respective 
space time dimensions. 
Boundary conditions  of all fields play a crucial role in the application of  localization in 
non-compact spaces.   Normalizable  boundary conditions are required for the definition of the 
path integral, it is only when normalizable boundary conditions are consistent with supersymmetric
boundary conditions that the method of localization can be applied. 
We have introduced a general set of assumptions on the second order operators that 
occur in the evaluation of the one loop determinants that hold
for theories with at least ${\cal N}=2$ supersymmetry. 
Under these assumptions we have constructed the Greens function and shown that 
the variation of the one loop determinant about the localizing background reduces to a total 
derivative. This is our first main result of the paper.
This  implies that the variation receives contributions only from asymptotic infinity and 
at the origin of $AdS$. 
Then from studying the asymptotics of the Greens function and the second order operators 
we show that the variation  of the one loop determinant is given by an integer times the 
variation of $\frac{1}{2} \ln Q^2$.  This is the second main result of our paper.

We then examine ${\cal N}=2$ Chern-Simons theory coupled to chiral multiplets on  $AdS_2\times S^1$
and show how the general set of assumptions we introduced hold for this case. 
We use our results to conclude that $U(N_c)$ Chern-Simons theory  at level $k$ coupled
to $N_f$ chiral multiplets and $N_f$ anti-chiral multiplets in the fundamental 
obeys Sieberg duality on $AdS_2\times S^1$.

As we have emphasised, the  Greens function method  is general as is applicable for other 
situations.  We believe that the method is applicable to evaluate one loop determinants that arise 
in localization of supersymmetric theories on $AdS_n\times S^m$ with at least ${\cal N}=2$ supersymmetry. 
One such case is that of ${\cal N}=2$ theories on $AdS_2\times S^2$ with matter. 
We hope to report results related to this in the near future. 
Localization of supersymmetric field theories  on $AdS_2\times S^2$ are relevant to 
evaluate quantum corrections to black hole entropy. 

Another direction to explore will be  localization of $2$-dimensional theories on $AdS_2$. 
In particular it will be interesting to see if the  duality between the Coulomb and the Higgs branch seen for 
 ${\cal N}=(2,2)$ theories  on the sphere $S^2$ by \cite{Doroud:2012xw} 
also hold for the case of the theory on $AdS_2$. 

The  general method we have introduced can be further refined. The 8 assumptions 
presented in section \ref{genproof} were obtained by 
 a detailed study of the Greens function approach and extracting general properties. 
 These assumptions enabled us to show that the variation of the one one loop determinant 
 reduces to a total derivative. We then introduced 3  assumptions in section \ref{calcbterm}. 
 These set of assumptions enabled us to show that the variation of the one loop determinant 
 is an integer times the variation of $\frac{1}{2} \ln Q^2$. 
Our preliminary investigations indicate that all these assumptions can  be shown to hold true  from the supersymmetry of the localizing Lagrangians. In fact we have seen that they also hold for ${\cal N}=2$ theories with matter
on $AdS_2\times S^2$ \cite{dggn}. 
It will be interesting to show that these assumptions follow 
as a natural  consequence of supersymmetry. 

  Finally, we have seen that the Greens function method shows that 
   variation of the one loop determinant is given by integer times the variation of $\frac{1}{2} \ln Q^2$. 
   We again  emphasise that this result is  only when normalizable boundary conditions are 
   compatible with supersymmetry. The integer is given by the index of the operator $C$ restricted to a $k$-dimensional vector space. It will be interesting to investigate if this result can be connected with the technique of applying 
   the fixed point evaluation of one loop determinants that arise in localization as recently applied 
   in \cite{Murthy:2015yfa, Assel:2016pgi,Cabo-Bizet:2017jsl, deWit:2018dix, Jeon:2018kec}.
   
\section{Acknowledgement}
We thank Sameer Murthy for useful conversations. The work of RG is supported by the ERC Consolidator Grant N. 681908, ``Quantum black holes: A macroscopic window into the microstructure of gravity''.
\appendix
\section{Supersymmetry of  the vector multiplet} \label{appenA}
Vector multiplet in $\mathcal N=2$ theory in Lorentzian signature contains a real scalar $\sigma$, gauge field $a_\mu$, an auxiliary real field $G$ and 2 component Weyl fermions $\lambda$ and $\tilde\lambda$. In order to compute partition function we need to analytically continue to Euclidean space. We choose the analytic continuation where the scalar field $\sigma$ and the auxiliary field $H$ are purely imaginary, the gauge field $a_\mu$ is real and the spinors $\lambda$ and $\tilde\lambda$ are two independent complex spinor.  
The Euclidean supersymmetry transformation of the fields in the vector multiplet is given by
\bea
&&Q_{s}\lambda=-\frac{i}{4}\epsilon \,H-\frac{i}{2}\epsilon^{\mu\nu\rho}\gamma_\rho F_{\mu\nu}\epsilon-i\gamma^\mu\epsilon\left(iD_\mu\sigma-V_\mu\sigma\right)\,,\nn\\
&&Q_{s}\tilde\lambda=\frac{i}{4}\tilde\epsilon \,H-\frac{i}{2}\epsilon^{\mu\nu\rho}\gamma_\rho F_{\mu\nu}\tilde\epsilon+i\gamma^\mu\tilde\epsilon\left(iD_\mu\sigma+V_\mu\sigma\right)\,,\nn\\
&&Q_{s}a_\mu=\frac{1}{2}\left(\epsilon\gamma_\mu\tilde\lambda+\tilde\epsilon\gamma_\mu\lambda\right)\,,\nn\\
&&Q_{s}\sigma=\frac{1}{2}\left(-\epsilon\tilde\lambda+\tilde\epsilon\lambda\right)\,,\nn\\
&&Q_{s}H=-2i\left[D_\mu\left(\epsilon\gamma^\mu\tilde\lambda-\tilde\epsilon\gamma^\mu\lambda\right)-i\left[\sigma,\epsilon\tilde\lambda+\tilde\epsilon\lambda\right]-iV_\mu\left(\epsilon\gamma^\mu\tilde\lambda+\tilde\epsilon\gamma^\mu\lambda\right)\right]\,.
\eea
The square of the susy transformations on vector multiplet fields are given by
\bea
&&Q_{s}^2\lambda\=\mathcal L_K\lambda+i[\Lambda,\lambda]-\frac{1}{2L}\lambda\,,\nn\\
&&Q_{s}^2\tilde\lambda\=\mathcal L_K\tilde\lambda+i[\Lambda,\tilde\lambda]+\frac{1}{2L}\tilde\lambda\,,\nn\\
&&Q_{s}^2a_\mu\=\mathcal L_K a_\mu+D_\mu\Lambda\,,\nn\\
&&Q_{s}^2\sigma\=\mathcal L_K\sigma-iK^\mu[a_\mu,\sigma]\,,\nn\\
&&Q_{s}^2H\=\mathcal L_K H+i[\Lambda,H]\,.
\eea
Here $\Lambda=\tilde\epsilon\epsilon\,\sigma-K^\rho a_\rho$. \\Using the above supersymmetry transformations we also note that $Q_{s}\Lambda\=0$.\\
Therefore, the algebra of supersymmetry transformation is given by
\be
Q_{s}^2\=\mathcal L_K+\delta^{\text{gauge transf}}_\Lambda+\delta^{R-\text{symm}}_{\frac{1}{2L}}\,.
\ee
\section{Equations of motions} \label{appenB}
The equations of motion for the vector field and scalar field can be written as
\be
M_{2}\p^{2}_{r}E^{+}(r)+M_{1}\p_{r}E^{+}(r)+M_{0}E^{+}(r)\=0
\ee
Here $M_{2,1,0}$ are $4\times 4$ matrices which are functions of coordinate $r$, and
\be
E^{+}(r)=\begin{pmatrix}a^{+}_{t}(r)\\a^{+}_{r}(r)\\a^{+}_{\th}(r)\\\sigma^{+}(r)\end{pmatrix}
\ee

 with components 
\be
M_{2}=\begin{pmatrix}-\frac{\sinh r}{2}&0&0&0\\0&\frac{\sinh r}{4L^{2}\xi}&0&0\\0&0&-\frac{1}{2L^{2}\sinh r}&0\\0&0&0&\frac{\sinh r}{2}\end{pmatrix}
\ee
\bea
&&M_{1,11}=-\frac{\cosh r}{2},\quad M_{1,12}=-\frac{1}{4\xi}\Big(n(1+2\xi)\cosh^{2}r-2\xi\a\Big)\sech r\,\tanh r\nn\\
&&M_{1,13}=0,\quad M_{1,14}=0\nn\\
&&M_{1,21}=\frac{1}{4\xi}\Big(n(1+2\xi)\cosh^{2}r-2\xi\a\Big)\sech r\,\tanh r,\quad M_{1,22}=\frac{\cosh r}{4L^{2}\xi}\nn\\
&&M_{1,23}=\frac{1}{4L^{2}\xi\sinh r\cosh^{2}r}\Big(p(1+2\xi)\cosh^{2}r-2L\xi\sinh^{2}r\Big),\quad M_{1,24}=0\nn\\
&&M_{1,31}=0,\quad M_{1,32}=-\frac{1}{4L^{2}\xi\sinh r\cosh^{2}r}\Big(p(1+2\xi)\cosh^{2}r-2L\xi\sinh^{2}r\Big)\nn\\
&&M_{1,33}=\frac{\cosh r}{2L^{2}\sinh^{2}r},\quad M_{1,34}=0,\quad M_{1,41}=0,\quad M_{1,42}=0,\quad M_{1,43}=0\nn\\
&&M_{1,44}=\frac{\cosh r}{2}.
\eea

\bea
&&M_{0,11}=\frac{p^{2}}{2\sinh r}+\frac{L^{2}}{2\xi}\Big(-n^{2}\sinh r+2\xi(2n-\a)\a\sech r\tanh r\Big)\nn\\
&&M_{0,12}=\a\sech^{3}r+\frac{1}{4\xi}\Big(n(1-2\xi)\cosh r-2(n+\xi\a)\sech r\Big)\nn\\
&&M_{0,13}=\frac{1}{2}(Ln-p)\a\sech r\tanh r-\frac{1}{4\xi\sinh r}\Big(p(n+2n\xi-2\a\,\xi)\Big),\nn\\
&&M_{0,14}=0,\quad M_{0,21}=\frac{n\sinh r\tanh r}{2\xi}\nn\\
&&M_{0,22}=\frac{1}{2}(n^{2}\sinh r-\a^{2}\sech r\tanh r)+\frac{1}{4L^{2}\xi\sinh r}\Big(-1+2p^{2}\xi+2\sinh^{2}r\tanh^{2}r\Big)\nn\\
&&M_{0,23}=-\frac{p}{L^{2}\xi\sinh r\sinh2r},\quad M_{0,24}=0\nn\\
&&M_{0,31}=\frac{1}{2}(Ln-p)\a\sech r\tanh r-\frac{1}{4\xi\sinh r}\Big(p(n+2n\xi-2\a\,\xi)\Big)\nn\\
&&M_{0,32}=\frac{1}{4L^{2}\xi}\Big(p(-1+2\xi)\frac{\cosh r}{\sinh^{2}r}+2\sech r(p-L\xi\a+2L\xi\a\sech^{2}r)\Big)\nn\\
&&M_{0,33}=\frac{1}{2L\sinh r}\Big(2p\a+L(n^{2}-\a^{2})\Big)-\frac{p^{2}}{4L^{2}\xi\sinh^{3}r}+\frac{\xi}{2L}(-2p+L\a)\sech r\tanh r,\nn\\
&&M_{0,34}=0,\quad M_{0,41}=0,\quad M_{0,42}=0,\quad M_{0,43}=0\nn\\
&&M_{0,44}=-\frac{p^{2}}{2\sinh r}-\frac{1}{2}L^{2}n^{2}\sinh r+\sech r\tanh r
\eea
In the case of fermions, we get
\bea
&&M_{2f,1i}=\delta_{i4}\,\frac{1}{2}\sech r\tanh r,\quad M_{2f,2i}=0,\quad M_{2f,3i}=\delta_{i4}\,\frac{1}{2L}\sech r\tanh r\nn\\
&&M_{2f,5i}=0,\quad M_{2f,6i}=0,\nn\\
&&M_{2f,41}=-\frac{1}{2}\sech r\tanh r,\quad M_{2f,42}=0,\quad M_{2f,43}=-\frac{1}{2L}\sech r\tanh r,\nn\\
&&M_{2f,44}=\frac{i}{2L}(p+L(n-\a))\sech r\tanh r,\quad M_{2f,45}=0,\quad M_{2f,46}=0
\eea
\bea
&&M_{1f,11}=0,\quad M_{1f,12}=-\frac{i}{2}\sech r\tanh r,\quad M_{1f,13}=0,\quad M_{1f,14}=\frac{1}{4}(3-\cosh2r)\sech^{3}r\nn\\
&&M_{1f,15}=0,\quad M_{1f,16}=-\frac{1}{2}L\tanh^{2}r,\quad M_{1f,21}=\frac{i}{2}\sech r\tanh r,\quad M_{1f,22}=0\nn\\
&&M_{1f,23}=\frac{i}{2L}\sech r\tanh r,\quad M_{1f,24}=0,\quad M_{1f,25}=-\frac{i}{2}\sinh r,\quad M_{1f,26}= 0,\nn\\
&&M_{1f,31}=0,\quad M_{1f,32}=-\frac{i}{2L}\sech r\tanh r,\quad M_{1f,33}=0,\quad M_{1f,34}=\frac{1}{4L}(3-\cosh2r)\sech^{3}r\nn\\
&&M_{1f,35}=0,\quad M_{1f,36}=\frac{1}{2}\sech^{2}r,\quad M_{1f,41}=\frac{1}{4\cosh^{3}r}(-3+\cosh2r),\quad M_{1f,42}=0\nn\\
&&M_{1f,43}=\frac{1}{4L\cosh^{3}r}(-3+\cosh2r),\quad M_{1f,44}=-\frac{i(p+L(n-\a))}{4L\cosh^{3}r}(-3+\cosh2r)\nn\\
&&M_{1f,45}=0,\quad M_{1f,46}=0,\quad M_{1f,51}=0,\quad M_{1f,52}=\frac{i}{2}\sinh r,\quad M_{1f,53}=0,\quad M_{1f,54}=0,\nn\\
&&M_{1f,55}=0,\quad M_{1f,56}=0,\quad M_{1f,61}=-\frac{L}{2}\tanh^{2}r,\quad M_{1f,62}=0,\quad M_{1f,63}=\frac{1}{2}\sech^{2}r\nn\\
&&M_{1f,64}=0,\quad M_{1f,65}=0,\quad M_{1f,66}=0
\eea
\bea
&&M_{0f,11}=-\frac{iL}{2}(-p+L(n-\a))\sech r\tanh r,\quad M_{0f,12}=\frac{i}{4\cosh^{3}r}(-3+\cosh2r)\nn\\
&&M_{0f,13}=-\frac{i}{2}\sech^{2}r(Ln\sinh r+\frac{p}{\sinh r}),\quad M_{0f,14}=-\frac{1}{2}\sech^{2}r(L^{2}n^{2}\sinh r+\frac{p^{2}}{\sinh r})\nn\\
&&M_{0f,15}=\frac{i}{2}L^{2}n\sinh r,\quad M_{0f,16}=-L\sech^{2}r\tanh r,\quad M_{0f,21}=0\nn\\
&&M_{0f,22}=\frac{i}{2L}(p+L(n+\a))\sech r\tanh r,\quad M_{0f,23}=0,\quad M_{0f,24}=0,\quad M_{0f,25}=-i\sinh r\tanh r,\nn\\
&&M_{0f,26}=\frac{1}{2}(Ln\sinh^{2}r-p)\sech^{2}r,\quad M_{0f,31}=-\frac{i}{2}\sech^{2}r(Ln\sinh r+\frac{p}{\sinh r}),\nn\\&&M_{0f,32}=\frac{i}{4L\cosh^{3}r}(-3+\cosh2r),\quad M_{0f,33}=\frac{i}{2L\sinh r\cosh^{2}r}(-p+L(n+\a)),\nn\\
&&M_{0f,34}=-\frac{1}{2L}\sech^{2}r(L^{2}n^{2}\sinh r+\frac{p^{2}}{\sinh r}),\quad M_{0f,35}=\frac{ip}{2\sinh r},\quad M_{0f,36}=-\sech^{2}r\tanh r\nn\\
&&M_{0f,41}=\frac{1}{2}\sech^{2}r(L^{2}n^{2}\sinh r+\frac{p^{2}}{\sinh r}),\quad M_{0f,42}=0,\quad M_{0f,43}=\frac{1}{2L}\sech^{2}r(L^{2}n^{2}\sinh r+\frac{p^{2}}{\sinh r})\nn\\
&& M_{0f,44}=-\frac{i(p+L(n-\a))}{2L\sinh r\cosh^{2}r}(L^{2}n^{2}\sinh^{2}r+p^{2}),\quad M_{0f,45}=0,\quad M_{0f,46}=0\nn\\
&&M_{0f,51}=\frac{i}{2}L^{2}n\sinh r,\quad M_{0f,52}=-\frac{i}{4\cosh r}(-3+\cosh2r),\quad M_{0f,53}=\frac{ip}{2\sinh r},\quad M_{0f,54}=0,\nn\\
&&M_{0f,55}=iL\xi(p+L(n-\a))\sinh r,\quad M_{0f,56}=0,\quad M_{0f,61}=0,\quad M_{0f,62}=-\frac{1}{2}(Ln\sinh^{2}r-p)\sech^{2}r,\nn\\
&&M_{0f,63}=0,\quad M_{0f,64}=0,\quad M_{0f,65}=0,\quad M_{0f,66}=-\frac{iL}{2}(p+L(n-\a))\sech r\tanh r
\eea

\section{On gauge fixing conditions} \label{appenC}
Here we justify the choice of the gauge fixing condition~\eqref{gfLagrangian}. In particular, we will show that the one loop result obtained for the abelian gauge theory in~\cite{David:2016onq} using the covariant gauge also holds true for the gauge fixing condition chosen in this paper. In fact, it works for the general gauge fixing condition
\begin{equation}
  G(a)\= \cosh^{\delta}r\nabla_{\m}(\frac{1}{\cosh^{\delta}r} a^{\m})=\cosh^{\delta}r\nabla_{\hat\m}(\frac{1}{\cosh^{\delta}r} a^{\hat\m})+ \partial_t a_t\,.
\end{equation}
where~$\hat\mu$ is 2-dim AdS indices.  The integral over ghost gives the Jacobian  $J$  which is defined through the 
functional integral as:
\begin{equation}
 \int {\cal{D}}\lambda\,  J\, \delta( M \lambda ) \= 1
 \label{J}
\end{equation}
where $M$ is obtained by infinitesimal gauge transformation $a \rightarrow a+ d\lambda$ on $G$:
\begin{equation}
   M \=\cosh^{\delta}r\nabla_{\hat\m}\Big(\frac{g^{\hat\m\hat\n}}{\cosh^{\delta}r} \nabla_{\hat\n}\Big)+ \partial_t^2
\end{equation}
and~$\lambda$ is in the space of all allowed gauge transformations. \\
Now, the allowed gauge transformations are defined as the ones that preserve the square integrability of gauge fields~$(a,a_t)$ : For~$t$-dependent part of the gauge transformation parameter~$\lambda(t,r,\theta)\=\sum_{n\neq0} \l_{n}(r,\theta)\,e^{int}$, it requires that 
\be
e^{r/2}\l_{n}(r,\theta)\rightarrow 0,\quad \text{for}\,\,r\rightarrow \infty\,.
\ee
The space of such gauge transformation we denote by~${\cal H}$.
However, for~$t$-independent part of the gauge transformation parameter~$\l_{0}(r,\theta)$, the condition on the normalizability of the gauge field requires that
\be
\l_{0}(r,\theta)\sim \l^{(0)}_{0}(\theta)+e^{-\beta\, r/2}\l^{(\beta)}_{0}(\theta)+...,\quad \text{for}\,\,\beta>1
\ee
The space of such gauge transformation we denote by~${\cal H}_{0}$.
We note that the operator $M$ is not self-adjoint for $\delta \neq 0$. However, it has the following properties: 

1) We first note that the differential operator $M$ does not have zero modes for $\delta>1$.  
We find that the solution of $Mf(r)=0$ has following large $r$ asymptotic
\begin{equation}\label{norm-f}
f(r) = 
\begin{cases} e^{-\frac{r}{2}(1-\delta-\sqrt{4L^{2}n^{2}+(1-\delta)^{2}})} , \qquad  \hbox{for}\; \; n\neq 0 \\
e^{r(\delta-1)}, \qquad  \hbox{for}\; \; n=0,\delta>1\,,\\
\mathcal O(1),\qquad \hbox{for}\; \; n=0,0\leq\delta\leq1
\end{cases}
\end{equation}
From the above we see that for $n\neq 0$, there are no zero mode. For $n=0$, we have zero modes for $\delta\leq 1$. Thus for $\delta=2$, which is our choice of gauge, we do not have zero modes and therefore, the gauge choice completely fixes the gauge. For $\delta=0$, which corresponds to covariant gauge fixing, there are an infinite number of zero modes~\cite{Banerjee:2010qc}.
Now, we will solve the differential equation for the adjoint operator, $M^{\dagger}\bar f(r)=0$. Solutions have following large $r$ asymptotic behaviour
\begin{equation}\label{norm-f}
\bar f(r) = 
\begin{cases} e^{-\frac{r}{2}(1+\delta-\sqrt{4L^{2}n^{2}+(1-\delta)^{2}})} , \qquad  \hbox{for}\; \; n\neq 0 \\
e^{-r}, \qquad  \hbox{for}\; \; n=0,\a>1\,,\\
e^{-\delta r},\qquad \hbox{for}\; \; n=0,0\leq\delta<1\\
\Gamma(0),\qquad \hbox{for}\; \; n=0,\delta=1
\end{cases}
\end{equation}
So, we see that for $n\neq 0$, there are no zero modes if $L^{2}n^{2}\geq \frac{2\delta-1}{4}$. In particular, for $\delta=2$, which is our gauge fixing, and $L^{2}>\frac{3}{4}$, which is supersymmetric case, there are no zero modes. 
For $0\leq\delta\leq 1$, we do not have zero modes.

2) For $n\neq 0$, $ M {\cal H}$ spans all of ${\cal H}$. The argument is as follows: if we assume that there must exist some function say $f'$ such that it is orthogonal to $M f$ for all 
$f \in {\cal H}$, ie. $\int d^2 x f' M f=0$ for all $f \in {\cal H}$, then integrating by part, we get $M^{\dagger} f'=0$. In this computation we obtain boundary terms which vanish because both $f,f' \in  {\cal H}$. 
But as we have shown before that for $n\neq 0$, kernel of $M^{\dagger}$  in ${\cal H}$ is empty. This proves that $M {\cal H}$ spans all of ${\cal H}$.

3) For $n=0$, $\lambda \in {\cal H}_0$. Now one can see that $M  {\cal H}_0$ is contained in  ${\cal H}$. Furthermore, $M$ has no kernel in $ {\cal H}_0$ for  $ \delta \geq 1$ while for $\delta < 1 $ it has a kernel with the zero mode going to order one  asymptotically. This means that for $\delta <1$ the large gauge transformations are not fixed.

We will now perform the path integral for each Fourier mode $n$ along $t$. For $n \neq 0$ we can solve $G(a)=0$ for $a_t$ as  
\begin{equation} a^{(n)}_t = \frac{i}{n} \cosh^{\delta}r \nabla_{\hat\m}\Big( \frac{g^{\hat\m\hat\n}}{\cosh^{\delta}r} a^{(n)}_{\hat\n} \Big)
\end{equation}
Integrating $a_t$ for $n\neq 0$, one also gets $\prod_{n\neq 0}\frac{1}{n}$. For $n=0$, we are left with the Gauge fixing condition $\cosh^{\delta}r \nabla_{\hat\m}\Big( \frac{g^{\hat\m\hat\n}}{\cosh^{\delta}r} a^{(0)}_{\hat\n} \Big)=0$. \\
For the other component of the gauge field, we use the 2-dim hodge decomposition  
\begin{equation}
a^{(n)}_{\hat\m}= \p_{\hat\m} f_{n} + \epsilon_{\hat\m\hat\n}\p^{\hat\n} f_{n}',\quad \text{where}\,\, \epsilon_{r\theta}=\sqrt{\hat g}=L^{2}\sinh r\,.
\end{equation}
Here $f' \in {\cal H}$  while $f \in {\cal H}_0$. Furthermore, we split $f$ as $\hat f+g$ 
where now $\hat f\in {\cal H}$ 
and $g \in {\cal H}_0/{\cal H} $, i.e. $g$ can go as $\mathcal O(1)$ near $r\rightarrow \infty$. Moreover, we demand that $g$ is orthogonal to all the normalizable fns $\hat f$ (and $f'$)
with respect to the inner product~$\int d^{2}x \,g^{\hat\m\hat\n}\p_{\hat\m}g \p_{\hat\n}\hat f=0$. This can be satisfied by taking $g$ to be solution of AdS laplacian (i.e.discreet modes). More explicitly a normalized basis with respect to the above norm for these solutions is~$g=\frac{1}{\sqrt{2 \pi |\ell|}}g_{n,p}\tanh^{|\ell|}(\frac{r}{2})e^{i\ell\theta}e^{in t}$, where~$\ell=\pm 1,\pm 2,....$.

 The Chern Simons action upto total derivative terms  becomes
 \begin{eqnarray}
 &&\frac{\kappa}{2} \int d^2 x \sum_{n>0}\Big[\frac{2 i}{n} \Bigl (( M \hat f_{n}) \,\Box f'_{-n} - ( M \hat f_{-n})\,\Box f'_{n} ]\Big)
 +.....\Big]+\frac{\kappa}{2} \int d^2 x a^{(0)}_{t} \,\Box f'_{0} \\
&& +\kappa \sum_{n,\ell>0} n(g_{n,-\ell}\, g_{-n,\ell}-g_{n,\ell}\, g_{-n,-\ell})\,. 
\label{action}
\end{eqnarray}
Here $\Box$ is a 2 AdS$_{2}$ Laplacian and dots include terms that do not involve $\hat f_n$ and $\hat f_{-n}$ and vanish when $f'_n$ and $f'_{-n}$ vanish.
Next, we want to change the variables from $f \rightarrow \wt{f} = M \hat f$. This can be done 
in the functional integral by inserting
\be
\int \prod_{n\neq 0}{\cal D} \wt{f}_n {\cal D} \wt{f}_{-n} \delta( \wt{f}_n-  M \hat f_n)\delta(\wt{f}_{-n}- M \hat f_{-n})\=1\,,
\ee
and integrate first $\hat f_n$ and $\hat f_{-n}$. Now, taking into account the Fadeev Popov determinant $J$ (for $n\neq0$) satisfying (\ref{J}) and noting the fact that $\hat f$ and $\lambda$ are in the same space of normalizable 
functions, the result of this integral together with $J$, is simply to replace $M \hat f_{n}$ and $M \hat f_{-n}$ by 
$\wt{f}_n$ and $\wt{f}_{-n}$, respectively in the action (\ref{action}). From the argument
in (2)  above $\wt{f}$ spans all of  ${\cal H}$. Now we can integrate $\wt{f}_n$ and $\wt{f}_{-n}$ and the result 
is $\prod_{n\neq 0}\delta( \frac{ i \kappa}{n} \Box f'_{n})\delta( \frac{- i \kappa}{n} \Box f'_{-n})$. Similarly, integrating the auxiliary fields $a^{(0)}_{t}$ we get $\delta( i \kappa \Box f'_{0})$. Since $f' \in {\cal H}$ and $\Box$ has no zero mode in ${\cal H}$, 
delta function enforces $f_{n}'=0$  in the remaining part of the action
i.e. the dots in (\ref{action}) vanish. Finally integrating $f'_n$ and $f'_{-n}$ we get $\frac{1}{\text{det}\kappa\, \Box}\prod_{n\neq 0}\frac{1}{\text{det} ( \frac{\kappa}{n}\, \Box)}$.\\
Now, we are still left with the integral over $f_{0}$ and $\delta(Mf_{0})$ in the integrand (coming from the gauge condition for $n=0$) where $f_{0} \in {\cal H}_0$ includes both square integrable as well as functions that go as order one at infinity. However, the contribution of this integral is exactly cancelled by the Fadeev Popov determinant $J$ for $n=0$.
Finally, we perform the integral over $g_{n,\ell}$ and take the logarithm of the entire result. We get
\begin{equation}
-\sum_{n\in\mathbb Z}\ln (k\,\Box)-\sum_{n,\ell=1}^{\infty}\ln k^{2}=-\frac{1}{2}\ln k
 \label{n}
\end{equation}
In the above we have used zeta function regularization. This is the same result as obtained in \cite{David:2016onq} using the covariant gauge fixing condition.



\begin{thebibliography}{10}
\bibitem{Pestun:2016zxk}
V.~Pestun et~al., {\it {Localization techniques in quantum field theories}},
  \href{http://xxx.lanl.gov/abs/1608.0295}{{\tt arXiv:1608.0295}}.

\bibitem{Witten:1992xu}
E.~Witten, {\it {Two-dimensional gauge theories revisited}},  {\em J. Geom.
  Phys.} {\bf 9} (1992) 303--368,
  [\href{http://xxx.lanl.gov/abs/hep-th/9204083}{{\tt hep-th/9204083}}].

\bibitem{Nekrasov:2003af}
N.~A. Nekrasov, {\it {Seiberg-Witten prepotential from instanton counting}},
  in {\em {International Congress of Mathematicians (ICM 2002) Beijing, China,
  August 20-28, 2002}}, 2003.
\newblock \href{http://xxx.lanl.gov/abs/hep-th/0306211}{{\tt hep-th/0306211}}.

\bibitem{Nekrasov:2003rj}
N.~Nekrasov and A.~Okounkov, {\it {Seiberg-Witten theory and random
  partitions}},  {\em Prog. Math.} {\bf 244} (2006) 525--596,
  [\href{http://xxx.lanl.gov/abs/hep-th/0306238}{{\tt hep-th/0306238}}].

\bibitem{Pestun:2007rz}
V.~Pestun, {\it {Localization of gauge theory on a four-sphere and
  supersymmetric Wilson loops}},  {\em Commun. Math. Phys.} {\bf 313} (2012)
  71--129, [\href{http://xxx.lanl.gov/abs/0712.2824}{{\tt arXiv:0712.2824}}].

\bibitem{Marino:2009jd}
M.~Marino and P.~Putrov, {\it {Exact Results in ABJM Theory from Topological
  Strings}},  {\em JHEP} {\bf 06} (2010) 011,
  [\href{http://xxx.lanl.gov/abs/0912.3074}{{\tt arXiv:0912.3074}}].

\bibitem{Benini:2016rke}
F.~Benini, K.~Hristov, and A.~Zaffaroni, {\it {Exact microstate counting for
  dyonic black holes in AdS4}},  {\em Phys. Lett.} {\bf B771} (2017) 462--466,
  [\href{http://xxx.lanl.gov/abs/1608.0729}{{\tt arXiv:1608.0729}}].

\bibitem{Benini:2015eyy}
F.~Benini, K.~Hristov, and A.~Zaffaroni, {\it {Black hole microstates in
  AdS$_{4}$ from supersymmetric localization}},  {\em JHEP} {\bf 05} (2016)
  054, [\href{http://xxx.lanl.gov/abs/1511.0408}{{\tt arXiv:1511.0408}}].

\bibitem{Dabholkar:2010uh}
A.~Dabholkar, J.~Gomes, and S.~Murthy, {\it {Quantum black holes, localization
  and the topological string}},  {\em JHEP} {\bf 06} (2011) 019,
  [\href{http://xxx.lanl.gov/abs/1012.0265}{{\tt arXiv:1012.0265}}].

\bibitem{Dabholkar:2011ec}
A.~Dabholkar, J.~Gomes, and S.~Murthy, {\it {Localization \& Exact
  Holography}},  {\em JHEP} {\bf 04} (2013) 062,
  [\href{http://xxx.lanl.gov/abs/1111.1161}{{\tt arXiv:1111.1161}}].

\bibitem{Gupta:2012cy}
R.~K. Gupta and S.~Murthy, {\it {All solutions of the localization equations
  for N=2 quantum black hole entropy}},  {\em JHEP} {\bf 02} (2013) 141,
  [\href{http://xxx.lanl.gov/abs/1208.6221}{{\tt arXiv:1208.6221}}].

\bibitem{Dabholkar:2014ema}
A.~Dabholkar, J.~Gomes, and S.~Murthy, {\it {Nonperturbative black hole entropy
  and Kloosterman sums}},  {\em JHEP} {\bf 03} (2015) 074,
  [\href{http://xxx.lanl.gov/abs/1404.0033}{{\tt arXiv:1404.0033}}].

\bibitem{Murthy:2015yfa}
S.~Murthy and V.~Reys, {\it {Functional determinants, index theorems, and exact
  quantum black hole entropy}},  {\em JHEP} {\bf 12} (2015) 028,
  [\href{http://xxx.lanl.gov/abs/1504.0140}{{\tt arXiv:1504.0140}}].

\bibitem{Gupta:2015gga}
R.~K. Gupta, Y.~Ito, and I.~Jeon, {\it {Supersymmetric Localization for BPS
  Black Hole Entropy: 1-loop Partition Function from Vector Multiplets}},  {\em
  JHEP} {\bf 11} (2015) 197, [\href{http://xxx.lanl.gov/abs/1504.0170}{{\tt
  arXiv:1504.0170}}].

\bibitem{Dabholkar:2014wpa}
A.~Dabholkar, N.~Drukker, and J.~Gomes, {\it {Localization in supergravity and
  quantum $AdS_4/CFT_3$ holography}},  {\em JHEP} {\bf 10} (2014) 90,
  [\href{http://xxx.lanl.gov/abs/1406.0505}{{\tt arXiv:1406.0505}}].

\bibitem{Cabo-Bizet:2017jsl}
A.~Cabo-Bizet, V.~I. Giraldo-Rivera, and L.~A. Pando~Zayas, {\it {Microstate
  counting of AdS$_{4}$ hyperbolic black hole entropy via the topologically
  twisted index}},  {\em JHEP} {\bf 08} (2017) 023,
  [\href{http://xxx.lanl.gov/abs/1701.0789}{{\tt arXiv:1701.0789}}].

\bibitem{David:2016onq}
J.~R. David, E.~Gava, R.~K. Gupta, and K.~Narain, {\it {Localization on
  AdS$_{2} \times$ S$^{1}$}},  {\em JHEP} {\bf 03} (2017) 050,
  [\href{http://xxx.lanl.gov/abs/1609.0744}{{\tt arXiv:1609.0744}}].

\bibitem{David:2018pex}
J.~R. David, E.~Gava, R.~K. Gupta, and K.~Narain, {\it {Boundary conditions and
  localization on AdS. Part I}},  {\em JHEP} {\bf 09} (2018) 063,
  [\href{http://xxx.lanl.gov/abs/1802.0042}{{\tt arXiv:1802.0042}}].

\bibitem{dggn}
J.~R. David, E.~Gava, R.~K. Gupta, and K.~Narain, {\it { Boundary conditions
  and localization on $AdS$: Part 3 }},  {\em To Appear}.

\bibitem{Doroud:2012xw}
N.~Doroud, J.~Gomis, B.~Le~Floch, and S.~Lee, {\it {Exact Results in D=2
  Supersymmetric Gauge Theories}},  {\em JHEP} {\bf 05} (2013) 093,
  [\href{http://xxx.lanl.gov/abs/1206.2606}{{\tt arXiv:1206.2606}}].

\bibitem{Assel:2016pgi}
B.~Assel, D.~Martelli, S.~Murthy, and D.~Yokoyama, {\it {Localization of
  supersymmetric field theories on non-compact hyperbolic three-manifolds}},
  {\em JHEP} {\bf 03} (2017) 095,
  [\href{http://xxx.lanl.gov/abs/1609.0807}{{\tt arXiv:1609.0807}}].

\bibitem{deWit:2018dix}
B.~de~Wit, S.~Murthy, and V.~Reys, {\it {BRST quantization and equivariant
  cohomology: localization with asymptotic boundaries}},  {\em JHEP} {\bf 09}
  (2018) 084, [\href{http://xxx.lanl.gov/abs/1806.0369}{{\tt
  arXiv:1806.0369}}].

\bibitem{Jeon:2018kec}
I.~Jeon and S.~Murthy, {\it {Twisting and localization in supergravity:
  equivariant cohomology of BPS black holes}},  {\em JHEP} {\bf 03} (2019) 140,
  [\href{http://xxx.lanl.gov/abs/1806.0447}{{\tt arXiv:1806.0447}}].

\bibitem{Banerjee:2010qc}
S.~Banerjee, R.~K. Gupta, and A.~Sen, {\it {Logarithmic Corrections to Extremal
  Black Hole Entropy from Quantum Entropy Function}},  {\em JHEP} {\bf 03}
  (2011) 147, [\href{http://xxx.lanl.gov/abs/1005.3044}{{\tt
  arXiv:1005.3044}}].

\end{thebibliography}
\end{document}